\documentclass[aps,prl,twocolumn,reprint, superscriptaddress]{revtex4-2}

\usepackage[caption=false]{subfig}
\usepackage{textgreek}
\usepackage{soul}
\usepackage{amsmath}
\usepackage{amssymb}
\usepackage{bm}
\usepackage{physics}
\usepackage{verbatim}
\usepackage[scr]{rsfso}
\usepackage{mwe}
\usepackage{graphicx}
\usepackage{overpic}
\usepackage[dvipsnames]{xcolor}
\definecolor{urlblue}{RGB}{6,69,173}

\usepackage[
	colorlinks=true,
	linkcolor=blue,
	anchorcolor=blue,
	citecolor=blue,
	urlcolor=blue,
	pdftitle={graviton_fci!}
]{hyperref}
\usepackage{tikz}
\usetikzlibrary{decorations.markings}

\tikzset{
  ->-/.style={
    decoration={markings, mark=at position #1 with {\arrow{latex}}},
    postaction={decorate}
  }
}

\renewcommand{\paragraph}[1]{\textit{#1}.---}

\makeatletter
\newcommand{\thetitle}{\@title}
\makeatother

\begin{document}

\title{Chiral Graviton Modes in Non-Abelian lattice Fractional Quantum Hall states}

\author{Zeno Bacciconi}
\thanks{These authors contribute equally}
\affiliation{SISSA --- International School for Advances Studies, via Bonomea 265, 34136 Trieste, Italy}
\affiliation{ICTP --- The Abdus Salam International Centre for Theoretical Physics, Strada Costiera 11, 34151 Trieste, Italy}

\author{Min Long}
\thanks{These authors contribute equally}
\affiliation{Department of Physics and HK Institute of Quantum Science \& Technology, The University of Hong Kong, Pokfulam Road,  Hong Kong SAR, China}
\affiliation{State Key Laboratory of Optical Quantum Materials, The University of Hong Kong, Pokfulam Road,  Hong Kong SAR, China}

\author{Hernan B. Xavier}
\affiliation{ICTP --- The Abdus Salam International Centre for Theoretical Physics, Strada Costiera 11, 34151 Trieste, Italy}
\affiliation{SISSA --- International School for Advances Studies, via Bonomea 265, 34136 Trieste, Italy}

\author{Hongyu Lu}
\affiliation{New Cornerstone Science Lab, Department of Physics, The University of Hong Kong, Pokfulam Road, Hong Kong SAR, China}
\affiliation{HK Institute of Quantum Science \& Technology, The University of Hong Kong, Pokfulam Road, Hong Kong SAR, China}
\affiliation{State Key Laboratory of Optical Quantum Materials, The University of Hong Kong, Pokfulam Road,  Hong Kong SAR, China}

\author{Marcello Dalmonte}
 \email{mdalmont@ictp.it}
\affiliation{ICTP --- The Abdus Salam International Centre for Theoretical Physics, Strada Costiera 11, 34151 Trieste, Italy}
\affiliation{Dipartimento di Fisica e Astronomia, Università di Bologna, and INFN, Sezione di Bologna, via Irnerio 46, I-40126 Bologna, Italy}

 \author{Zi Yang Meng}
 \email{zymeng@hku.hk}
\affiliation{Department of Physics and HK Institute of Quantum Science \& Technology, The University of Hong Kong, Pokfulam Road,  Hong Kong SAR, China}
\affiliation{State Key Laboratory of Optical Quantum Materials, The University of Hong Kong, Pokfulam Road,  Hong Kong SAR, China}

\date{\today}

\begin{abstract} 

Synthetic quantum matter provides a highly tunable route to fractional quantum Hall physics beyond the constraints of conventional electronic materials. However, previous theoretical studies have mostly focused on their ground state properties. It remains unclear to what extent such platforms could reveal key excitation properties of fractional quantum Hall states. Here, we study charge-neutral collective excitations in a non-Abelian lattice fractional quantum Hall state realized in the bosonic Harper–Hofstadter model at unity filling factor ($\nu=1$), realizing a Moore–Read ground state. 
Combining full exact diagonalization, band-projected exact diagonalization, and matrix-product-state simulations, we demonstrate the existence of a long-lived chiral graviton mode, probed by chiral 3-body correlators, for the first time on lattice non-Abelian states. 
The graviton signal is topological sector-independent and could be observed via geometric quenches in small open droplets directly relevant to current cold-atom experiments, while other neutral modes, such as the magnetoroton and neutral fermion, are less resolved at presently achievable volumes.

\end{abstract}
\maketitle
\paragraph{\textcolor{blue}{\it Introduction}}
The fractional quantum Hall (FQH) effect is a macroscopic quantum effect that emerges in 2D electron gases from the combination of a perpendicular magnetic field and Coulomb interaction. In particular, the Moore-Read state provides the first example of non-Abelian topological order and has been believed to be a candidate for the $\nu = 5/2$ state~\cite{moore1991nonabelions,Willett1987observation,Greiter1992Paired}. The potential to realize topological quantum computing has spurred sustained efforts to explore such phases across a variety of experimental settings. In electronic systems, tunneling~\cite{Dolev2008observation,Radu2008quasiparticle}, interferometric~\cite{Willett2013magnetic,Willett2023interference,Kim2026ABinterference}, and thermal transport measurements~\cite{Banerjee2018observation} have reported observations compatible with non-abelian topological order.

Besides electronic systems, synthetic quantum systems have emerged as promising platforms for {\it bosonic} FQH physics due to their high tunability~\cite{goldman2016topological}.  The laser-assisted-tunneling realizations of the Harper–Hofstadter (HH) Hamiltonian~\cite{jaksch2003creation} for ultracold atoms in optical lattices~\cite{aidelsburgerRealization2013,miyake2013realizing} have made these systems candidates for correlated Hall physics. Along this path, experiments have demonstrated edge mode dynamics in synthetic ribbons~\cite{stuhl2015visualizing,mancini2015observation}, Meissner-like effects~\cite{atala2014observation,impertro2024realization}, correlations in single tweezer droplets~\cite{lunt2024realization}, and, very recently, observed evidence of p-wave pairing for the Pfaffian state, where $3$-body correlations have been measured~\cite{Joyce2026Pfaffian}. 

On the theoretical front, the ground-state properties of the HH Hamiltonian have been extensively studied. Matrix Product State (MPS)~\cite{palm2021pfaffian,Boesl_prb2022_topomott} simulations indicate that the bosonic Harper–Hofstadter (HH)/Bose–Hubbard model can stabilize a $\nu=1$ Moore–Read phase at low flux per plaquette $\phi \lesssim 0.2$. A natural question arises: can cold atom platforms provide further insight, which goes beyond the ground-state properties, on the excitation content of FQH liquids \cite{binatiRepellin_prr2024_spectroscopyFQH,nardin2026hallviscositymetricsensitivedichroic,nardin2024quantum} - in particular, at currently available volumes, that are limited by noise? An intriguing question is the properties of the chiral graviton mode (CGM) on such system~\cite{gromov2017bimetric,liu2018quench,liou2019chiral,nguyen2022multiple,kumar2022neutral,yang2016acoustic,liuResolving2024,Balram_2024,Yang_prl2012_modelwavefunctions_graviton,Nguyen2021Dirac,Wang2022Analytic,wang2023Geometric,Haldane2021Graviton,pu2024microscopic,bacciconi_prx2025_gravitonpolaritons,AjitPapic_prx2022_highpartons}. The CGM carries angular momentum $2$ and is a genuine feature of FQH states arising from fluctuation of their intrinsic metric. Another important issue is the fate of related charge-neutral modes on this platform, such as the magneto-roton~\cite{gmp1986magnetoroton} and the neutral fermion~\cite{Moller2011neutral,Bonderson2011Numerical,Repellin2015projective}.

\begin{figure*}[htp!]
    \centering
    \begin{overpic}[width=1\linewidth]{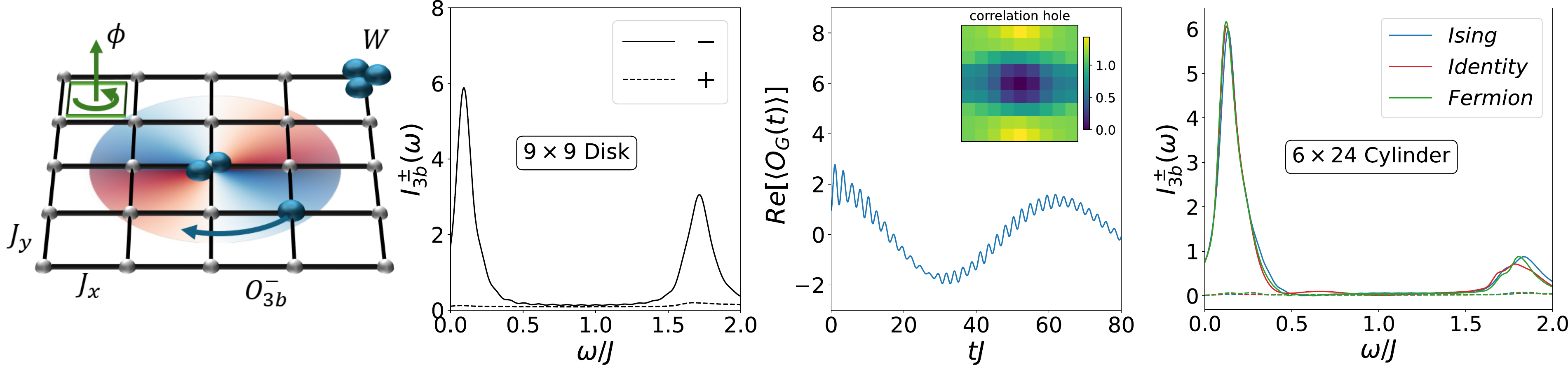}
    
     \put(0,20){(a)}
     \put(31,20){(b)}
     \put(55,20){(c)}
     \put(80,20){(d)}
     
    \end{overpic}
    \caption{\textbf{Pfaffian state graviton mode and geometric quench}. (a) Sketch of the interacting bosonic Harper-Hofstadter system and the graviton operator $O^-_{3b}$ used to track the dynamics of the graviton mode. 
    (b) Graviton spectrum from MPS simulations on an open disk geometry (flux $\phi=1/6$, $W=\infty$ and $N=5$ particles). 
    (c) Real-time dynamics of graviton displacement $\langle O_{G}(t) \rangle/\langle O_{G}(0) \rangle $ after a geometric quench (anisotropy $\alpha=0.8\to1.$). Same parameters as (b) Inset: 3-body correlation $C^{3}_{ij}$ in Eq.~\eqref{eq:eq3} measured at $\eta = 0.8$. 
    (d) Graviton spectrum from MPS simulations in cylinder geometry, for different topological sectors (in different colors) and chirality (in solid or dashed lines); the system size is $6\times 24$, and the same flux and 3-body hard-core condition as those in (b). 
    }
    \label{fig:disksimu}
\end{figure*}
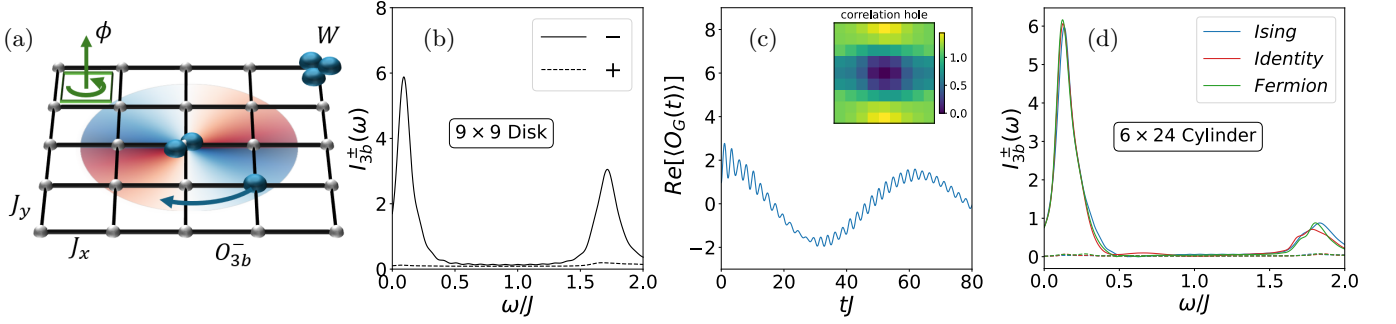
In this work we address the above question by studying the $\nu=1$ Moore–Read phase at low fluxes for the bosonic HH model. Combining full exact diagonalization (fED), projected ED (pED), and Matrix Product State(MPS) simulations, we report the existence of a lattice CGM, probed by chiral 3-body correlators; hence establishing a strong connection between CGM and correlation-hole dynamics. We show that the CGM signal is robust among disk, cylinder, and torus geometries. Through disk simulations, we show that such graviton mode is clearly observable even for a $9 \times 9$ disk filled with $N = 5$ particles, well within experimental reach~\cite{impertro2024realization,Joyce2026Pfaffian}. Interestingly, although the CGM is a long-wavelength excitation, its spectrum measured from 3 different topological sectors of $\nu =1$ bosonic Moore-Read state shows similar features. By a quantitative lifetime analysis ~\cite{longChiral2026}, we show that the CGM in the bosonic MR state is a long-lived excitation in the thermodynamic limit with a relative decay rate $\Gamma_G/\omega_G\sim 0.1$. We then further study how finite momentum neutral modes of non-abelian states, magnetoroton and neutral fermion mode, are more susceptible to lattice effects and can be distinguished only for very low fluxes. 

\paragraph{\textcolor{blue}{\it Model and Method}} We study the bosonic HH model on a square lattice, 
\begin{equation}
H=-\sum_{\langle i,j\rangle_a;a \in{x,y}}J_{a} (b^\dagger_ib_{j}  e^{i\phi_{ij}}+h.c.) 
+ \frac{W}{6}\sum_i (b^\dagger_i)^3 (b_i)^3 
\end{equation}
with the generic anisotropic tunneling rates $J_x = J$ and $J_y = \alpha J$ illustrated in Fig.~\ref{fig:disksimu}(a); $\langle i,j\rangle_{x/y}$ are nearest neighbors on a specific direction and $W$ is a three-body on site interaction. We set $J=1$ as the energy unit, and the phases $\phi_{ij}$ realize a HH model at flux per plaquette $\phi$. We note that, in most cold atom experiments, there is an effective $W=\infty$ interaction due to three body losses and Zeno dynamics~\cite{daley2009atomic}. We thus work on a local projected Hilbert space in our MPS simulations, and only use finite $W$ in pED. While three body interactions provide a model Hamiltonian for Moore-Read states, similar results are expected for Hubbard-type interactions \cite{regnault_prl2003_bosonsV0,Boesl_prb2022_topomott,palm2021pfaffian}.

Following previous constructions of graviton operators in abelian lattice FQH states~\cite{longSpectra2025,longChiral2026}, we introduce a discrete version of the 3-body continuum stress tensor \cite{liou2019chiral,Haldane2021Graviton} as the operator to probe the CGM:
\begin{equation}
    \begin{aligned}
    O^{\pm}_{3b} & =\sum_{r_i,\delta} e^{\pm 2i \arg{[\delta]}}f_G(\delta) n_{r_i}(n_{r_i}-1) n_{r_i+\delta}, 
    \end{aligned}
\label{eq:definition_O3n}
\end{equation}
where $\arg{[\delta]}$ is the angle given by the vector $\delta = \vec{r}_i-\vec{r}_j$ and $f_G(\delta)$ a short ranged function (here $f(1)=1$, $f(\sqrt{2})=1/\sqrt{2}$ and 0 otherwise). Importantly when the flux is low enough $\phi\to0$ and the continuum limit is approached, the lattice graviton operator of Eq.~\eqref{eq:definition_O3n} exactly flows to that of a continuum model system with three body interactions (see the derivation in Supplementary Material (SM)~\cite{suppmat}). The contribution in $O^{\pm}_{3b}$ from a single site is schematically plot in Fig.~\ref{fig:disksimu}(a).

We study this Hamiltonian using fED, pED, and MPS simulations~\cite{White1992,White1993,Schollwock2011,Haegeman2011Time,Haegeman2016Unifying} and the spectral density of the graviton operator:
\begin{align}\label{eq:definition_I3n}
  I^\pm_{3b}(\omega)=\sum_{n}|\bra{n} O^\pm_{3b}\ket{0}|^2 \delta(\omega-E_n+E_0)
\end{align}
where $\ket{n}$ are many-body eigenstates and $E_n$ their energies (also see SM~\cite{suppmat} for the method of spectrum calculation and convergence check). In the MPS simulation, we study both cylinder and disk geometries. The ground state is determined with bond dimension up to $D = 2400$ and truncation error $\epsilon<10^{-8}$. In the time evolution, we keep the $D = 800$ state with a truncation error $\epsilon<10^{-6}$.

\paragraph{\textcolor{blue}{\it Chiral Graviton Mode on disk and cylinder}} Chiral graviton modes, i.e., spin-2 collective excitations of FQH liquids, have been recently shown to be robust excitations for abelian fractional quantum hall states on the lattice~\cite{longSpectra2025,xavier2025chiralgravitonslattice,longChiral2026}. The presence and characterization of lattice CGM on more fragile non-abelian states is the open challenge we tackle here.

We first focus on a finite FQH droplet, which is realized in an $L_x\times L_y$ square with open edges along both $x$ and $y$ directions and is relevant for current cold atom platforms ~\cite{leonard2023realization,Joyce2026Pfaffian}. As a starting point, we consider a $9\times 9$ disk filled by $N = 5$ particles and with flux $1/6$. The number of particles $N$ should be chosen such that the classical radius of the FQH droplet $R = \sqrt{\frac{N}{2 \pi \nu \phi }} < \frac{L_x}{2}$, thus the disk edge has negligible effect on the dynamics of the FQH droplet. In disk geometry, the particle density near the edges is low. Therefore, the entanglement brought by the outer sites is irrelevant, and we can access larger $L_y$ compared with cylinder geometry. Figure~\ref {fig:disksimu}(b) shows the graviton spectrum measured in a $9\times9$ size under the 3-body hard-core condition $W = \infty$ (a very small trapping potential $V(r) = V_0 r^2$ is added to trap the droplet, here $V_0 = 2\times10^{-4}$ and $r$ is the distance measured from center of the disk, for the ground state properties, see SM~\cite{suppmat}). The energy of CGM locates around  $\omega \sim 0.09$ with chirality clearly resolved, indicating that such excitations remain robust and detectable even in a small lattice droplet. Higher energy features around $\omega\sim 1.5$ can be attributed to inter-band transitions (i.e. they are not present in band-projected calculations).

Having observed the CGM signal of the Pfaffian state on a disk, a natural follow-up question is, does the graviton spectrum depend on the topological sector? To address this question, we put the Hamiltonian on a $L_y = 6$ cylinder and perform an infinite DMRG (iDMRG) study~\cite{White1993}, where we can determine the topological sector by computing the momentum-resolved entanglement spectrum~\cite{Pollmann2012detection,Cincio2013Characterizing} (also see SM~\cite{suppmat} for the detailed result). We employ a hybrid approach where the time evolution is performed on an $6\times 24$ MPS segment embedded in two infinite half chains to eliminate finite-size effects~\cite{Phien2012infinite,Milsted2013Variational}. The result is shown in Fig.~\ref{fig:disksimu}(d). The spectrum does not depend on the topological sector with energy $\omega_G \sim 0.13$, which is slightly higher than that of the disk case, which we attribute to finite size effects.

\paragraph{\textcolor{blue}{\it Geometric quench on disk}} While graviton-modes have been measured by Raman scattering in 2DEG~\cite{liang2024evidence,Yang2026_natphys_gravitonpartons}, geometric quenches, that is a sudden change in the mass anisotropy $g^{ij}$, have also been proposed to access the intrinsic metric properties of FQH liquids~\cite{liu2018quench,liu2021quench,ippoliti2018geometry}. Such geometric quench in the lattice model can be realized as a sudden change of the tunneling ratios $\eta = J_x/J_y$~\cite{xavier2025chiralgravitonslattice} and are very natural for such platforms.
In particular, quenching $\eta$ could be achieved by suddenly changing the lattice depth along $x$, or by modulating $J_{x/y}$ in time via tuning the two-photon Rabi frequency related to photon assisted tunneling~\cite{jaksch2003creation,impertro2024realization}. 

We first test the robustness of the FQH state to ensure the system stays in the same phase throughout the quench protocol. The ground-state fidelity $F(\eta) = \langle \psi(\eta) | \psi(\eta + \delta \eta) \rangle$ is measured and is close to 1, as shown in SM~\cite{suppmat}, indicating the FQH state is stable in the parameter window $\eta \in [0.8,1]$. Having prepared the anisotropic ground state at $\eta = 0.8$, we quench the system Hamiltonian at $\eta=1$ and start the time evolution. 

In the inset of Fig.~\ref{fig:disksimu} (c), we show the 3-body correlation holes $C^3_{ij}$: 
\begin{equation}\label{eq:eq3}
    C^3_{ij}=\frac{\langle n_i(n_i-1) n_j\rangle  -2\delta_{ij}(\langle n_i n_j\rangle + \langle n_i\rangle)}{\langle n_i(n_i-1)\rangle \langle n_j\rangle}.
\end{equation}
of the anisotropic ground state at $\eta=0.8$. The $C^3_{ij}$ clearly reveal the anisotropy of the ground state, indicating the close connection between intrinsic metric and correlation hole shape. We then monitor the evolution of graviton displacement operators $O_G = O^+ + O^-$, which can be reconstructed from three-body correlations \cite{Joyce2026Pfaffian}, and show the result in the main panel of Fig.~\ref{fig:disksimu}(c). Two oscillation modes could be read from $\langle O_G(t) \rangle$, corresponding to a low-frequency peak (graviton mode) and a high-frequency peak (inter-band transitions) appearing in the spectrum. Furthermore, from the rotation of the squeezed correlation hole, the chirality of the graviton mode can be seen (SM~\cite{suppmat}). These results remarkably shows that, even in a small-sized disk containing only a few particles and with open boundary, experimental studies of bulk graviton modes are possible, provided that timescales of order of a few tenths of $J^{-1}$ can be reached.


\paragraph{\textcolor{blue}{\it Intrinsic lifetime analysis}} The graviton-mode is very often not the lowest energy bulk neutral excitation of an FQH liquid \cite{gmp1986magnetoroton,LiuXiang_prb2024_geometricexcitations}. It is then important to understand whether it actually constitutes a well-defined quasi-particle with long-lifetime, or the available decay channels involving scattering into pairs of lower-energy modes turn it in a broad continuum feature when the system size is large enough~\cite{wangDynamics2025}.

\begin{figure}
    \centering
    \begin{overpic}
[width=\linewidth]{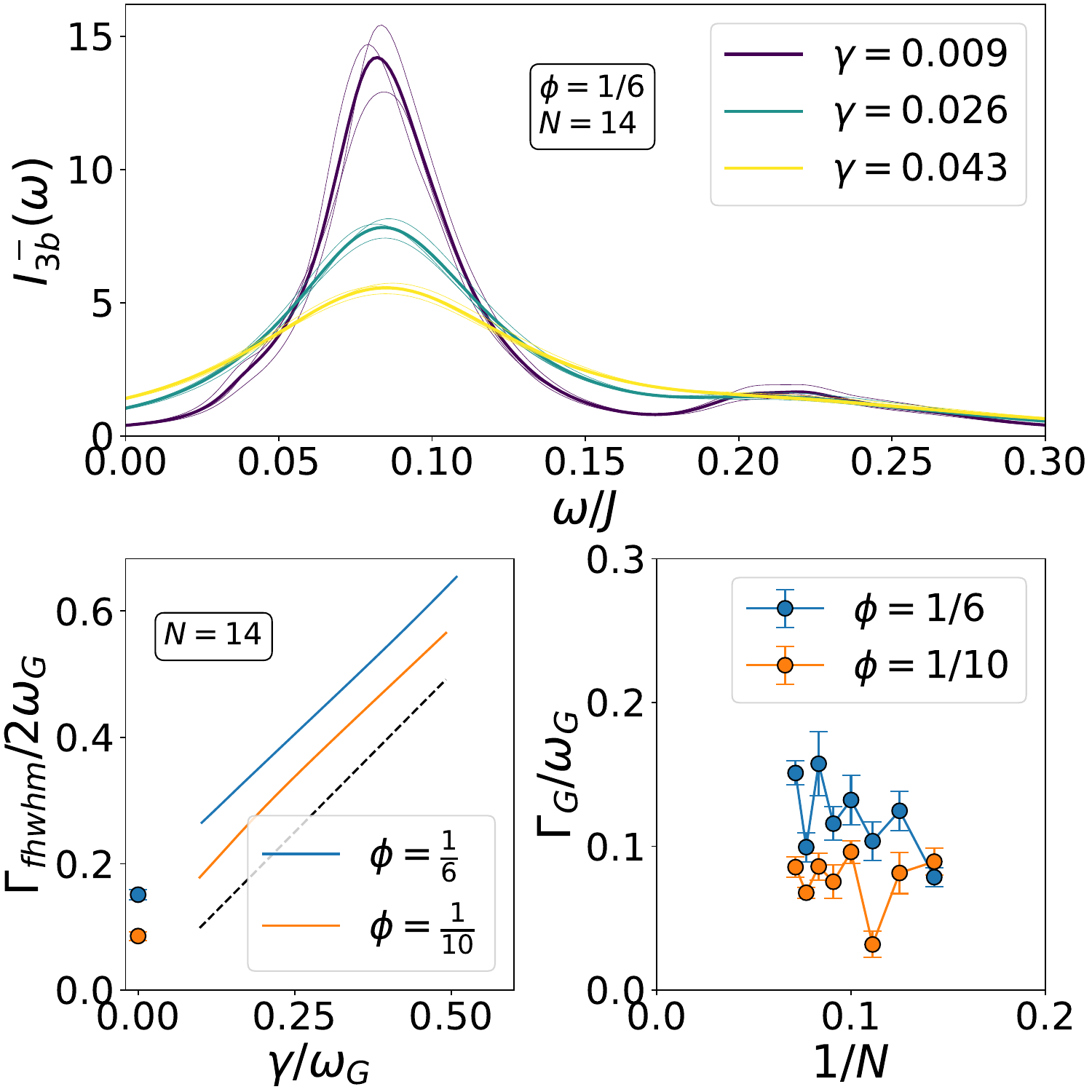}
        \put(13,93){(a)}
        \put(13,35){(b)}
        \put(62,35){(c)}

    \end{overpic}
    \caption{\textbf{Life time analysis on the Graviton mode} (a) Graviton spectra at different regularization parameters $\gamma$. Different ground states are shown as thin lines, while the thicker line correspond to the averaged spectra. (b) Rescaled Full width half maxima $\Gamma_{fwhm}$ vs. $\gamma$ for different fluxes $\phi$. The dashed black line represents the infinite lifetime scenario $\Gamma_{fwhm}(\gamma)=2\gamma$, while the $\gamma=0$ points are extrapolations. (c) Graviton intrinsic decay rate $\Gamma_G$ relative to its energy $\omega_G$ as a function of $\phi$ and particle number. All results are obtained with pED at $W=5$}
    \label{fig:lifetime}
\end{figure}

In continuum Landau Levels, experiments on abelian states have reported the existence of well-resolved spectral features \cite{liang2024evidence} with long lifetimes. Available numerical spectra also indicate narrow spectral features, despite the frequency of graviton modes lying in a two-particle continuum for both abelian \cite{liou2019chiral} and non-abelian \cite{Haldane2021Graviton} states. On lattice FQH and FCI abelian states, long graviton lifetimes have been recently reported \cite{longSpectra2025,longChiral2026}. In the following, we will perform a similar analysis to reveal the intrinsic graviton lifetime for the non-Abelian states under study.

In order to access different system sizes, needed for a faithful extraction of the graviton lifetime, we perform exact diagonalization projected on the lowest band of a torus geometry at different fluxes $n_\phi$, whose results are shown in Fig. \ref{fig:lifetime}. This is approximate for a finite interaction $W$, but captures the main effects of scattering into many-body excitations which predominantly live in the lowest band as shown in SM~\cite{suppmat}. The spectral functions defined in Eq. \eqref{eq:definition_I3n} are then obtained with Lanczos continued fraction methods \cite{Weisse_rmp2008_kpm,ED_spectralfunctions_koch} and importantly depend on the regularization parameter $\gamma$. As the target mode lives around a continuum of states, the regularization $\gamma$ should be larger than the many-body spacing $\gamma \gtrsim \delta$ around the graviton energy $\omega_G$ so to not be dominated by finite size features. On the other hand this shall not be too large to loose resolution on the graviton mode itself $\gamma \lesssim \omega_G$. 

In Fig.~\ref{fig:lifetime} (a) we show an example graviton spectral function for a system at flux $n_\phi=1/6$ and $N=14$ particles ($7\times14$ torus) at different regularization $0.1\omega_G\lesssim \gamma\lesssim0.5\omega_G$. Results from different ground states are shown as thin lines while the thicker lines are ground-states-averaged results. As $\gamma$ is reduced, a clear peak emerges. The full-width-half-maximum of this peak $\omega\sim 0.9$ is tracked as a function of $\gamma$ in Fig.~\ref{fig:lifetime}(b) for different fluxes. The behavior $\Gamma_{fwhm}(\gamma)$ can be understood in terms of a quasiparticle with a finite imaginary self-energy whose Green's function can be expressed around the resonant frequency as $G_{O_{3b}^-}(\omega+i\gamma)\simeq\frac{1}{\omega+i\gamma-\omega_G +i\Gamma_G}$. Since $I_{3b}^-(\omega)=-\frac{1}{\pi}\Im G_{O_{3b}^-}(\omega)$, we expect a lorentzian shape with:
\begin{align}\label{eq:Gamma_est}
    \Gamma_{fwhm}=2(\Gamma_G+\gamma)
\end{align}
as shown in Fig.~\ref{fig:lifetime} (a). The actual value of $\Gamma_G$ is then estimated using Eq.~\eqref{eq:Gamma_est}, averaged over the shown range of $\gamma$. An error bar is then associated to this corresponding to the standard deviation of the estimate in the proposed range $\gamma\in[0.1\omega_G,0.5\omega_G]$  where $\omega_G$ correspond to the peak position. 

The estimate for the intrinsic decay rate $\Gamma_G$ is shown at $\gamma=0$ in Fig.~\ref{fig:lifetime}(b) and as a function of system size in Fig.~\ref{fig:lifetime}(c). The latter shows a strong even-odd effect which prevents a proper $1/N$ extrapolation. However, as oscillations are limited in amplitude, these data provide strong evidence for a finite decay rate in the thermodynamic limit. The value of such decay rate decreases by reducing the flux per plaquette $n_\phi$, although the overall value seems larger ($\Gamma_G/\omega_G\sim 0.08$) than what is found for abelian states at similar fluxes ($\Gamma_G^{\nu=1/3}/\omega_G\sim 0.02$)~\cite{longChiral2026}. Overall, this agrees with qualitatively less resolved spectra in continuum LL \cite{liou2019chiral,Haldane2021Graviton} for Moore-Read states compared to Laughlin states.

\begin{figure}
    \centering
    \begin{overpic}[width=1.\linewidth]{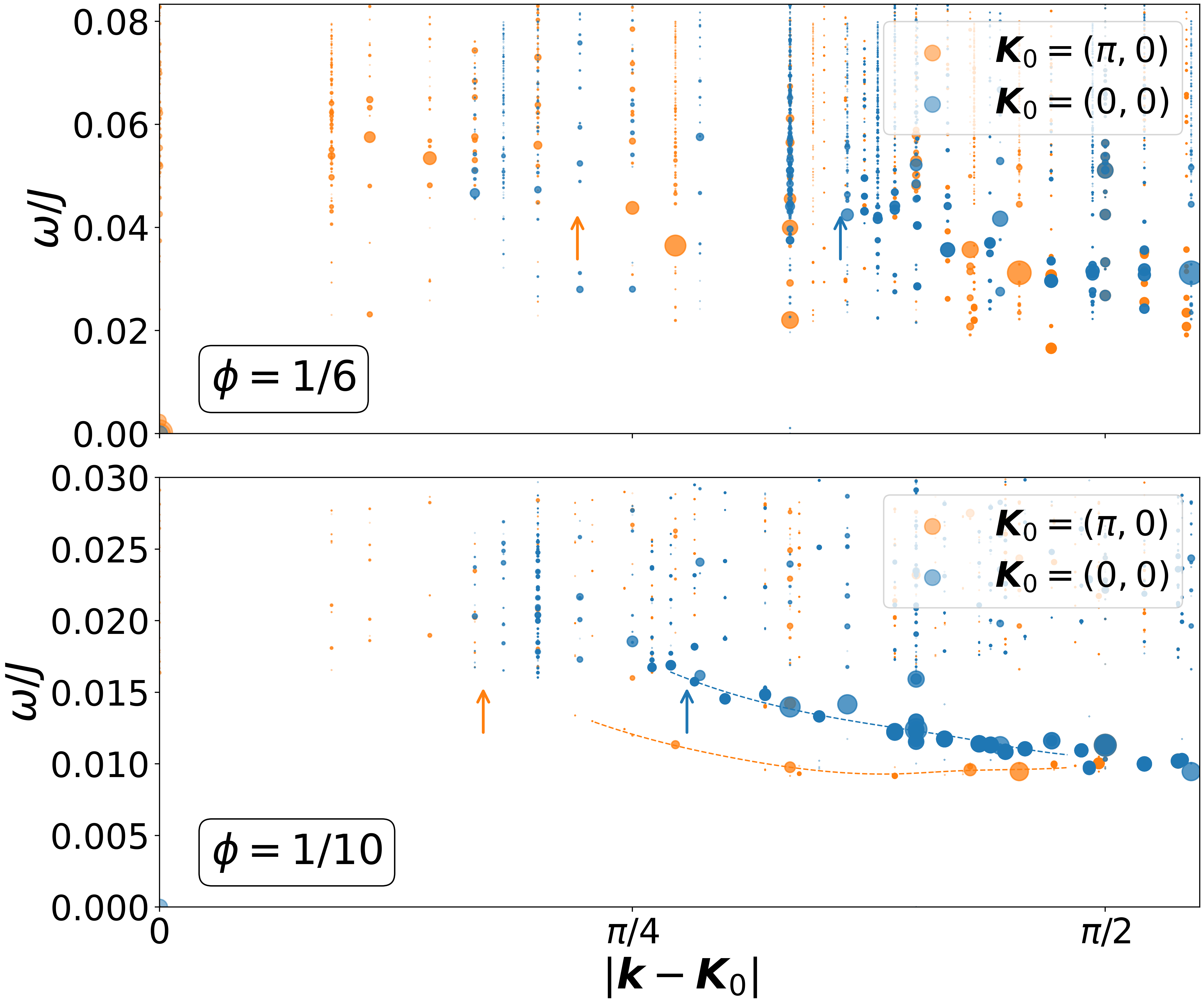}
    \put(15,78){(a)}
    \put(15,38){(b)}
    \end{overpic}
    \caption{\textbf{Magneto-roton mode and neutral fermion mode.} Neutral modes dispersion for a \textit{high} flux $n_\phi=1/6$ (panel (a)) and a \textit{low} flux $n_\phi=1/10$ (panel (b)) obtained with pED. Points correspond to eigenstates $\ket{n}$ with a weight (marker area) corresponding to $|\bra{n}n_{\boldsymbol{k}}\ket{0}|$. As explained in the text, the relative momenta $|\boldsymbol{k}-\boldsymbol{K}_0|$ reveal the magnetoroton mode (blue) (all sizes $N=6,..,13$) and the neutral fermion mode (orange and only odd sizes). Arrows indicate where the corresponding continuum mode enters the two-particle continuum; dashed lines are a guide to the eye.}
    \label{fig:neutral_modes}
\end{figure}

\paragraph{\textcolor{blue}{\it Magneto-roton and Neutral Fermion Excitations}}
Magnetorotons are finite-momentum charge-neutral excitations of both FQH and FCI systems~\cite{gmp1986magnetoroton,Hongyu_prl2024_thermodinamicFCI} and are typically the lowest energy bulk states above the ground state manifold. They control density modulations on top of the uniform liquid and, for simple abelian states, can be understood as bound states of quasi-particles and quasi-holes. For non-abelian states, however, other neutral modes are expected at low energy. Moore-Read states in the continuum feature both a magnetoroton and a neutral fermion mode~\cite{Yang_prl2012_modelwavefunctions_graviton,CooperMoller_prl2011_neutralfermion,Repellin2015projective}. 

The two neutral modes can be distinguished in continuum systems by the fact that they are centered around different high-symmetry points in the center of mass momenta Brillouin zone \cite{Repellin2015projective}. In particular, whenever a ground state is present on these points $\boldsymbol{K}_0$, the dispersion is of magnetoroton character while if no ground state is present there is a neutral fermion dispersion. On the lattice the exact center of mass symmetry is lost and its Brillouin zone is mapped on the smaller many-body lattice momenta Brillouin zone \cite{Bernevig2012Emergent}. Nevertheless, for torus geometries with size $N\times q$ where $n_\phi=1/q$, two of the four relevant momentum points $\boldsymbol{K}_0=(0,0)$ and $\boldsymbol{K}_0=(\pi,0)$ are well embedded into the lattice Brillouin zone (see SM \cite{suppmat} for details) and the dispersion around those can be studied on the lattice as well. 

To reveal the two dispersion relations (see Fig.~\ref{fig:neutral_modes}) we use matrix elements of density operator $|\bra{n} n_{\boldsymbol{k}}\ket{0}|$ with $n_{\boldsymbol{k}}=(L_xL_y)^{-1/2} \sum_i e^{i\boldsymbol{k}\cdot \boldsymbol{r}_i}n_i$. Here $\ket{0}$ is the ground state in the momentum sector $(0,0)$, and $\boldsymbol{k}\in\{(2\pi n_x/ L_x,2\pi n_y/L_y) ; n_{x/y}=0,..,L_{x/y}-1\}$ lives in an unfolded Brillouin zone. Hence each eigenstate $\ket{n}$ will actually show up in $q$ different unfolded momentum sectors. Crucially, by showing these matrix elements for different system sizes $N=6,...,13$ as a function of the resulting momentum distance $|\boldsymbol{k}-\boldsymbol{K}_0|$, both neutral modes are revealed. Note that only odd $N$ sizes feature the neutral fermion mode around $\boldsymbol{K}_0=(\pi,0)$ \cite{Repellin2015projective} and thus these are only shown when $\boldsymbol{K}_0=(\pi,0)$. 

Interestingly, the dispersions are clear only for very low fluxes $\phi=1/10$ (Fig. \ref{fig:neutral_modes}(b)), losing resolution when the flux is increased to $\phi=1/6$ (Fig. \ref{fig:neutral_modes}(a)). This apparently contrast with the well resolved graviton mode at the same flux (see Fig. \ref{fig:lifetime}); we remark that the lattice is expected to provide stronger perturbations at finite momenta $\boldsymbol{k}$, especially when $\boldsymbol{k}$ become comparable to the reciprocal lattice vectors. As a reference for this, we track the momenta where the modes enters the two particle continuum in a off-lattice model \cite{Repellin2015projective} (arrows in Fig. \ref{fig:neutral_modes}). These are inversely proportional $\boldsymbol{\kappa}_{cont.}\sim 1/l_B$ to the effective magnetic length $l_B= \sqrt{1/2\pi n_\phi }$ (units of lattice constant).

\paragraph{\textcolor{blue}{\it Discussion}} 
In this work, we have demonstrated that chiral graviton modes persist as robust collective excitations in non-Abelian lattice fractional quantum Hall states. Focusing on the bosonic Harper–Hofstadter model in the ($\nu=1$) Moore–Read regime, we introduced a chiral 3-body lattice operator that recovers the expected continuum graviton metric tensor in the continuum limit. Using a combination of fED, pED and MPS simulations, we showed that this operator resolves a clear chirality-selective graviton response across disk, cylinder, and torus geometries.

In terms of experimental impact, the key result of our study is that the graviton signal remains visible even in small open droplets containing only a few particles, detectable by a simple quench protocol. 
Our lifetime analysis indicates that the graviton mode remains long-lived despite lattice effects. Combined, our results show the remarkable promise cold atom experiments hold for illustrating features of FQH states that are very challenging for electronic systems - taking a somewhat alternative turn in the field, that has so far focused on reproducing established features of FQH already observed in solid state devices.

Beyond the graviton response, we identified finite-momentum magnetoroton excitations and neutral-fermion modes. Our results imply that, to go beyond ground-state studies and access excitation properties, the CGM, a long-wavelength excitation, appears to be experimentally accessible with current capabilities. In contrast, resolving the finite-wavelength dispersions of other neutral excitations, such as the magnetoroton and neutral fermion, remains more challenging.

Looking forward, it would be interesting to investigate whether the proposed neutral fermionic partner of the graviton mode, i.e. the \textit{gravitino} \cite{Yang_prl2012_modelwavefunctions_graviton,PuPapic_prl2023_moorereadsupersymmetry}, can also be studied on open lattice geometries.

\paragraph{\it Acknowledgement} We thank Ha Quang Trung, Bo Yang, Yuzhu Wang, Dung Nguyen Xuan, Ajit Balram, Cecile Repellin, Alberto Nardin, Iacopo Carusotto, Minh Pham and Dam Thanh Son for discussions on excitations in FQH systems. ML and ZYM acknowledge the support from the Research Grants Council (RGC) of Hong Kong (Project Nos. HKU C7037-22GF, 17302223, 17301924, 17301725), the ANR/RGC Joint Research Scheme sponsored by RGC of Hong Kong and French National Research Agency (Project No. A\_HKU703/22). ML thanks ICTP for its kind hospitality through the Project "Wave-function Networks: Probe and understand quantum many-body systems via network and complexity theory - WaveNets", funded by the European Union (Grant Agreement n. 101087692). ZB thanks the Department of Physics at the University of Hong Kong for kind hospitality, during which this work is carried out. M.~D. was partly supported by the EU-Flagship programme Pasquans2, by the PNRR MUR project PE0000023-NQSTI, and by the ERC Consolidator grant WaveNets (Grant agreement ID: 101087692). 
We thank HPC2021 system under the Information Technology Services at the University of Hong Kong~\cite{hpc2021}, as well as the Beijing Paratera Tech Corp., Ltd~\cite{paratera} for providing HPC resources that have contributed to the research results reported within this paper. 

\bibliographystyle{longapsrev4-2}
\bibliography{non_abelian_graviton}

\begin{thebibliography}{77}%
\makeatletter
\providecommand \@ifxundefined [1]{%
 \@ifx{#1\undefined}
}%
\providecommand \@ifnum [1]{%
 \ifnum #1\expandafter \@firstoftwo
 \else \expandafter \@secondoftwo
 \fi
}%
\providecommand \@ifx [1]{%
 \ifx #1\expandafter \@firstoftwo
 \else \expandafter \@secondoftwo
 \fi
}%
\providecommand \natexlab [1]{#1}%
\providecommand \enquote  [1]{``#1''}%
\providecommand \bibnamefont  [1]{#1}%
\providecommand \bibfnamefont [1]{#1}%
\providecommand \citenamefont [1]{#1}%
\providecommand \href@noop [0]{\@secondoftwo}%
\providecommand \href [0]{\begingroup \@sanitize@url \@href}%
\providecommand \@href[1]{\@@startlink{#1}\@@href}%
\providecommand \@@href[1]{\endgroup#1\@@endlink}%
\providecommand \@sanitize@url [0]{\catcode `\\12\catcode `\$12\catcode `\&12\catcode `\#12\catcode `\^12\catcode `\_12\catcode `\%12\relax}%
\providecommand \@@startlink[1]{}%
\providecommand \@@endlink[0]{}%
\providecommand \url  [0]{\begingroup\@sanitize@url \@url }%
\providecommand \@url [1]{\endgroup\@href {#1}{\urlprefix }}%
\providecommand \urlprefix  [0]{URL }%
\providecommand \Eprint [0]{\href }%
\providecommand \doibase [0]{https://doi.org/}%
\providecommand \selectlanguage [0]{\@gobble}%
\providecommand \bibinfo  [0]{\@secondoftwo}%
\providecommand \bibfield  [0]{\@secondoftwo}%
\providecommand \translation [1]{[#1]}%
\providecommand \BibitemOpen [0]{}%
\providecommand \bibitemStop [0]{}%
\providecommand \bibitemNoStop [0]{.\EOS\space}%
\providecommand \EOS [0]{\spacefactor3000\relax}%
\providecommand \BibitemShut  [1]{\csname bibitem#1\endcsname}%
\let\auto@bib@innerbib\@empty
\bibitem [{\citenamefont {Moore}\ and\ \citenamefont {Read}(1991)}]{moore1991nonabelions}%
  \BibitemOpen
  \bibfield  {author} {\bibinfo {author} {\bibfnamefont {G.}~\bibnamefont {Moore}}\ and\ \bibinfo {author} {\bibfnamefont {N.}~\bibnamefont {Read}},\ }\bibfield  {title} {\bibinfo {title} {{Nonabelions in the fractional quantum Hall effect}},\ }\href {https://doi.org/10.1016/0550-3213(91)90407-O} {\bibfield  {journal} {\bibinfo  {journal} {Nucl. Phys. B}\ }\textbf {\bibinfo {volume} {360}},\ \bibinfo {pages} {362} (\bibinfo {year} {1991})}\BibitemShut {NoStop}%
\bibitem [{\citenamefont {Willett}\ \emph {et~al.}(1987)\citenamefont {Willett}, \citenamefont {Eisenstein}, \citenamefont {St\"{o}rmer}, \citenamefont {Tsui}, \citenamefont {Gossard},\ and\ \citenamefont {English}}]{Willett1987observation}%
  \BibitemOpen
  \bibfield  {author} {\bibinfo {author} {\bibfnamefont {R.}~\bibnamefont {Willett}}, \bibinfo {author} {\bibfnamefont {J.~P.}\ \bibnamefont {Eisenstein}}, \bibinfo {author} {\bibfnamefont {H.~L.}\ \bibnamefont {St\"{o}rmer}}, \bibinfo {author} {\bibfnamefont {D.~C.}\ \bibnamefont {Tsui}}, \bibinfo {author} {\bibfnamefont {A.~C.}\ \bibnamefont {Gossard}},\ and\ \bibinfo {author} {\bibfnamefont {J.~H.}\ \bibnamefont {English}},\ }\bibfield  {title} {\bibinfo {title} {Observation of an even-denominator quantum number in the fractional quantum Hall effect},\ }\href {https://doi.org/10.1103/physrevlett.59.1776} {\bibfield  {journal} {\bibinfo  {journal} {Physical Review Letters}\ }\textbf {\bibinfo {volume} {59}},\ \bibinfo {pages} {1776–1779} (\bibinfo {year} {1987})}\BibitemShut {NoStop}%
\bibitem [{\citenamefont {Greiter}\ \emph {et~al.}(1992)\citenamefont {Greiter}, \citenamefont {Wen},\ and\ \citenamefont {Wilczek}}]{Greiter1992Paired}%
  \BibitemOpen
  \bibfield  {author} {\bibinfo {author} {\bibfnamefont {M.}~\bibnamefont {Greiter}}, \bibinfo {author} {\bibfnamefont {X.}~\bibnamefont {Wen}},\ and\ \bibinfo {author} {\bibfnamefont {F.}~\bibnamefont {Wilczek}},\ }\bibfield  {title} {\bibinfo {title} {Paired Hall states},\ }\href {https://doi.org/10.1016/0550-3213(92)90401-v} {\bibfield  {journal} {\bibinfo  {journal} {Nuclear Physics B}\ }\textbf {\bibinfo {volume} {374}},\ \bibinfo {pages} {567–614} (\bibinfo {year} {1992})}\BibitemShut {NoStop}%
\bibitem [{\citenamefont {Dolev}\ \emph {et~al.}(2008)\citenamefont {Dolev}, \citenamefont {Heiblum}, \citenamefont {Umansky}, \citenamefont {Stern},\ and\ \citenamefont {Mahalu}}]{Dolev2008observation}%
  \BibitemOpen
  \bibfield  {author} {\bibinfo {author} {\bibfnamefont {M.}~\bibnamefont {Dolev}}, \bibinfo {author} {\bibfnamefont {M.}~\bibnamefont {Heiblum}}, \bibinfo {author} {\bibfnamefont {V.}~\bibnamefont {Umansky}}, \bibinfo {author} {\bibfnamefont {A.}~\bibnamefont {Stern}},\ and\ \bibinfo {author} {\bibfnamefont {D.}~\bibnamefont {Mahalu}},\ }\bibfield  {title} {\bibinfo {title} {Observation of a quarter of an electron charge at the ν = 5/2 quantum Hall state},\ }\href {https://doi.org/10.1038/nature06855} {\bibfield  {journal} {\bibinfo  {journal} {Nature}\ }\textbf {\bibinfo {volume} {452}},\ \bibinfo {pages} {829–834} (\bibinfo {year} {2008})}\BibitemShut {NoStop}%
\bibitem [{\citenamefont {Radu}\ \emph {et~al.}(2008)\citenamefont {Radu}, \citenamefont {Miller}, \citenamefont {Marcus}, \citenamefont {Kastner}, \citenamefont {Pfeiffer},\ and\ \citenamefont {West}}]{Radu2008quasiparticle}%
  \BibitemOpen
  \bibfield  {author} {\bibinfo {author} {\bibfnamefont {I.~P.}\ \bibnamefont {Radu}}, \bibinfo {author} {\bibfnamefont {J.~B.}\ \bibnamefont {Miller}}, \bibinfo {author} {\bibfnamefont {C.~M.}\ \bibnamefont {Marcus}}, \bibinfo {author} {\bibfnamefont {M.~A.}\ \bibnamefont {Kastner}}, \bibinfo {author} {\bibfnamefont {L.~N.}\ \bibnamefont {Pfeiffer}},\ and\ \bibinfo {author} {\bibfnamefont {K.~W.}\ \bibnamefont {West}},\ }\bibfield  {title} {\bibinfo {title} {Quasi-Particle Properties from Tunneling in the $\nu= 5/2$ Fractional Quantum Hall State},\ }\href {https://doi.org/10.1126/science.1157560} {\bibfield  {journal} {\bibinfo  {journal} {Science}\ }\textbf {\bibinfo {volume} {320}},\ \bibinfo {pages} {899–902} (\bibinfo {year} {2008})}\BibitemShut {NoStop}%
\bibitem [{\citenamefont {Willett}\ \emph {et~al.}(2013)\citenamefont {Willett}, \citenamefont {Nayak}, \citenamefont {Shtengel}, \citenamefont {Pfeiffer},\ and\ \citenamefont {West}}]{Willett2013magnetic}%
  \BibitemOpen
  \bibfield  {author} {\bibinfo {author} {\bibfnamefont {R.~L.}\ \bibnamefont {Willett}}, \bibinfo {author} {\bibfnamefont {C.}~\bibnamefont {Nayak}}, \bibinfo {author} {\bibfnamefont {K.}~\bibnamefont {Shtengel}}, \bibinfo {author} {\bibfnamefont {L.~N.}\ \bibnamefont {Pfeiffer}},\ and\ \bibinfo {author} {\bibfnamefont {K.~W.}\ \bibnamefont {West}},\ }\bibfield  {title} {\bibinfo {title} {Magnetic-Field-Tuned Aharonov-Bohm Oscillations and Evidence for Non-Abelian Anyons at $\nu=5/2$},\ }\bibfield  {journal} {\bibinfo  {journal} {Physical Review Letters}\ }\textbf {\bibinfo {volume} {111}},\ \href {https://doi.org/10.1103/physrevlett.111.186401} {10.1103/physrevlett.111.186401} (\bibinfo {year} {2013})\BibitemShut {NoStop}%
\bibitem [{\citenamefont {Willett}\ \emph {et~al.}(2023)\citenamefont {Willett}, \citenamefont {Shtengel}, \citenamefont {Nayak}, \citenamefont {Pfeiffer}, \citenamefont {Chung}, \citenamefont {Peabody}, \citenamefont {Baldwin},\ and\ \citenamefont {West}}]{Willett2023interference}%
  \BibitemOpen
  \bibfield  {author} {\bibinfo {author} {\bibfnamefont {R.}~\bibnamefont {Willett}}, \bibinfo {author} {\bibfnamefont {K.}~\bibnamefont {Shtengel}}, \bibinfo {author} {\bibfnamefont {C.}~\bibnamefont {Nayak}}, \bibinfo {author} {\bibfnamefont {L.}~\bibnamefont {Pfeiffer}}, \bibinfo {author} {\bibfnamefont {Y.}~\bibnamefont {Chung}}, \bibinfo {author} {\bibfnamefont {M.}~\bibnamefont {Peabody}}, \bibinfo {author} {\bibfnamefont {K.}~\bibnamefont {Baldwin}},\ and\ \bibinfo {author} {\bibfnamefont {K.}~\bibnamefont {West}},\ }\bibfield  {title} {\bibinfo {title} {Interference Measurements of Non-Abelian $e/4$ Abelian $e/2$ Quasiparticle Braiding},\ }\bibfield  {journal} {\bibinfo  {journal} {Physical Review X}\ }\textbf {\bibinfo {volume} {13}},\ \href {https://doi.org/10.1103/physrevx.13.011028} {10.1103/physrevx.13.011028} (\bibinfo {year} {2023})\BibitemShut {NoStop}%
\bibitem [{\citenamefont {Kim}\ \emph {et~al.}(2026)\citenamefont {Kim}, \citenamefont {Dev}, \citenamefont {Shaer}, \citenamefont {Kumar}, \citenamefont {Ilin}, \citenamefont {Haug}, \citenamefont {Iskoz}, \citenamefont {Watanabe}, \citenamefont {Taniguchi}, \citenamefont {Mross}, \citenamefont {Stern},\ and\ \citenamefont {Ronen}}]{Kim2026ABinterference}%
  \BibitemOpen
  \bibfield  {author} {\bibinfo {author} {\bibfnamefont {J.}~\bibnamefont {Kim}}, \bibinfo {author} {\bibfnamefont {H.}~\bibnamefont {Dev}}, \bibinfo {author} {\bibfnamefont {A.}~\bibnamefont {Shaer}}, \bibinfo {author} {\bibfnamefont {R.}~\bibnamefont {Kumar}}, \bibinfo {author} {\bibfnamefont {A.}~\bibnamefont {Ilin}}, \bibinfo {author} {\bibfnamefont {A.}~\bibnamefont {Haug}}, \bibinfo {author} {\bibfnamefont {S.}~\bibnamefont {Iskoz}}, \bibinfo {author} {\bibfnamefont {K.}~\bibnamefont {Watanabe}}, \bibinfo {author} {\bibfnamefont {T.}~\bibnamefont {Taniguchi}}, \bibinfo {author} {\bibfnamefont {D.~F.}\ \bibnamefont {Mross}}, \bibinfo {author} {\bibfnamefont {A.}~\bibnamefont {Stern}},\ and\ \bibinfo {author} {\bibfnamefont {Y.}~\bibnamefont {Ronen}},\ }\bibfield  {title} {\bibinfo {title} {Aharonov–Bohm interference in even-denominator fractional quantum Hall states},\ }\href {https://doi.org/10.1038/s41586-025-09891-2} {\bibfield  {journal} {\bibinfo  {journal} {Nature}\ }\textbf {\bibinfo {volume}
  {649}},\ \bibinfo {pages} {323–329} (\bibinfo {year} {2026})}\BibitemShut {NoStop}%
\bibitem [{\citenamefont {Banerjee}\ \emph {et~al.}(2018)\citenamefont {Banerjee}, \citenamefont {Heiblum}, \citenamefont {Umansky}, \citenamefont {Feldman}, \citenamefont {Oreg},\ and\ \citenamefont {Stern}}]{Banerjee2018observation}%
  \BibitemOpen
  \bibfield  {author} {\bibinfo {author} {\bibfnamefont {M.}~\bibnamefont {Banerjee}}, \bibinfo {author} {\bibfnamefont {M.}~\bibnamefont {Heiblum}}, \bibinfo {author} {\bibfnamefont {V.}~\bibnamefont {Umansky}}, \bibinfo {author} {\bibfnamefont {D.~E.}\ \bibnamefont {Feldman}}, \bibinfo {author} {\bibfnamefont {Y.}~\bibnamefont {Oreg}},\ and\ \bibinfo {author} {\bibfnamefont {A.}~\bibnamefont {Stern}},\ }\bibfield  {title} {\bibinfo {title} {Observation of half-integer thermal Hall conductance},\ }\href {https://doi.org/10.1038/s41586-018-0184-1} {\bibfield  {journal} {\bibinfo  {journal} {Nature}\ }\textbf {\bibinfo {volume} {559}},\ \bibinfo {pages} {205–210} (\bibinfo {year} {2018})}\BibitemShut {NoStop}%
\bibitem [{\citenamefont {Goldman}\ \emph {et~al.}(2016)\citenamefont {Goldman}, \citenamefont {Budich},\ and\ \citenamefont {Zoller}}]{goldman2016topological}%
  \BibitemOpen
  \bibfield  {author} {\bibinfo {author} {\bibfnamefont {N.}~\bibnamefont {Goldman}}, \bibinfo {author} {\bibfnamefont {J.~C.}\ \bibnamefont {Budich}},\ and\ \bibinfo {author} {\bibfnamefont {P.}~\bibnamefont {Zoller}},\ }\bibfield  {title} {\bibinfo {title} {Topological quantum matter with ultracold gases in optical lattices},\ }\href@noop {} {\bibfield  {journal} {\bibinfo  {journal} {Nature Physics}\ }\textbf {\bibinfo {volume} {12}},\ \bibinfo {pages} {639} (\bibinfo {year} {2016})}\BibitemShut {NoStop}%
\bibitem [{\citenamefont {Jaksch}\ and\ \citenamefont {Zoller}(2003)}]{jaksch2003creation}%
  \BibitemOpen
  \bibfield  {author} {\bibinfo {author} {\bibfnamefont {D.}~\bibnamefont {Jaksch}}\ and\ \bibinfo {author} {\bibfnamefont {P.}~\bibnamefont {Zoller}},\ }\bibfield  {title} {\bibinfo {title} {{Creation of effective magnetic fields in optical lattices: the Hofstadter butterfly forcold neutral atoms}},\ }\href {https://doi.org/10.1088/1367-2630/5/1/356} {\bibfield  {journal} {\bibinfo  {journal} {New J. Phys.}\ }\textbf {\bibinfo {volume} {5}},\ \bibinfo {pages} {56} (\bibinfo {year} {2003})}\BibitemShut {NoStop}%
\bibitem [{\citenamefont {Aidelsburger}\ \emph {et~al.}(2013)\citenamefont {Aidelsburger}, \citenamefont {Atala}, \citenamefont {Lohse}, \citenamefont {Barreiro}, \citenamefont {Paredes},\ and\ \citenamefont {Bloch}}]{aidelsburgerRealization2013}%
  \BibitemOpen
  \bibfield  {author} {\bibinfo {author} {\bibfnamefont {M.}~\bibnamefont {Aidelsburger}}, \bibinfo {author} {\bibfnamefont {M.}~\bibnamefont {Atala}}, \bibinfo {author} {\bibfnamefont {M.}~\bibnamefont {Lohse}}, \bibinfo {author} {\bibfnamefont {J.~T.}\ \bibnamefont {Barreiro}}, \bibinfo {author} {\bibfnamefont {B.}~\bibnamefont {Paredes}},\ and\ \bibinfo {author} {\bibfnamefont {I.}~\bibnamefont {Bloch}},\ }\bibfield  {title} {\bibinfo {title} {Realization of the Hofstadter Hamiltonian with Ultracold Atoms in Optical Lattices},\ }\href {https://doi.org/10.1103/PhysRevLett.111.185301} {\bibfield  {journal} {\bibinfo  {journal} {Phys. Rev. Lett.}\ }\textbf {\bibinfo {volume} {111}},\ \bibinfo {pages} {185301} (\bibinfo {year} {2013})}\BibitemShut {NoStop}%
\bibitem [{\citenamefont {Miyake}\ \emph {et~al.}(2013)\citenamefont {Miyake}, \citenamefont {Siviloglou}, \citenamefont {Kennedy}, \citenamefont {Burton},\ and\ \citenamefont {Ketterle}}]{miyake2013realizing}%
  \BibitemOpen
  \bibfield  {author} {\bibinfo {author} {\bibfnamefont {H.}~\bibnamefont {Miyake}}, \bibinfo {author} {\bibfnamefont {G.~A.}\ \bibnamefont {Siviloglou}}, \bibinfo {author} {\bibfnamefont {C.~J.}\ \bibnamefont {Kennedy}}, \bibinfo {author} {\bibfnamefont {W.~C.}\ \bibnamefont {Burton}},\ and\ \bibinfo {author} {\bibfnamefont {W.}~\bibnamefont {Ketterle}},\ }\bibfield  {title} {\bibinfo {title} {{Realizing the Harper Hamiltonian with Laser-Assisted Tunneling in Optical Lattices}},\ }\href {https://doi.org/10.1103/PhysRevLett.111.185302} {\bibfield  {journal} {\bibinfo  {journal} {Phys. Rev. Lett.}\ }\textbf {\bibinfo {volume} {111}},\ \bibinfo {pages} {185302} (\bibinfo {year} {2013})}\BibitemShut {NoStop}%
\bibitem [{\citenamefont {Stuhl}\ \emph {et~al.}(2015)\citenamefont {Stuhl}, \citenamefont {Lu}, \citenamefont {Aycock}, \citenamefont {Genkina},\ and\ \citenamefont {Spielman}}]{stuhl2015visualizing}%
  \BibitemOpen
  \bibfield  {author} {\bibinfo {author} {\bibfnamefont {B.~K.}\ \bibnamefont {Stuhl}}, \bibinfo {author} {\bibfnamefont {H.-I.}\ \bibnamefont {Lu}}, \bibinfo {author} {\bibfnamefont {L.~M.}\ \bibnamefont {Aycock}}, \bibinfo {author} {\bibfnamefont {D.}~\bibnamefont {Genkina}},\ and\ \bibinfo {author} {\bibfnamefont {I.~B.}\ \bibnamefont {Spielman}},\ }\bibfield  {title} {\bibinfo {title} {{Visualizing edge states with an atomic Bose gas in the quantum Hall regime}},\ }\href {https://doi.org/10.1126/science.aaa8515} {\bibfield  {journal} {\bibinfo  {journal} {Science}\ }\textbf {\bibinfo {volume} {349}},\ \bibinfo {pages} {1514} (\bibinfo {year} {2015})}\BibitemShut {NoStop}%
\bibitem [{\citenamefont {Mancini}\ \emph {et~al.}(2015)\citenamefont {Mancini}, \citenamefont {Pagano}, \citenamefont {Cappellini}, \citenamefont {Livi}, \citenamefont {Rider}, \citenamefont {Catani}, \citenamefont {Sias}, \citenamefont {Zoller}, \citenamefont {Inguscio}, \citenamefont {Dalmonte},\ and\ \citenamefont {Fallani}}]{mancini2015observation}%
  \BibitemOpen
  \bibfield  {author} {\bibinfo {author} {\bibfnamefont {M.}~\bibnamefont {Mancini}}, \bibinfo {author} {\bibfnamefont {G.}~\bibnamefont {Pagano}}, \bibinfo {author} {\bibfnamefont {G.}~\bibnamefont {Cappellini}}, \bibinfo {author} {\bibfnamefont {L.}~\bibnamefont {Livi}}, \bibinfo {author} {\bibfnamefont {M.}~\bibnamefont {Rider}}, \bibinfo {author} {\bibfnamefont {J.}~\bibnamefont {Catani}}, \bibinfo {author} {\bibfnamefont {C.}~\bibnamefont {Sias}}, \bibinfo {author} {\bibfnamefont {P.}~\bibnamefont {Zoller}}, \bibinfo {author} {\bibfnamefont {M.}~\bibnamefont {Inguscio}}, \bibinfo {author} {\bibfnamefont {M.}~\bibnamefont {Dalmonte}},\ and\ \bibinfo {author} {\bibfnamefont {L.}~\bibnamefont {Fallani}},\ }\bibfield  {title} {\bibinfo {title} {{Observation of chiral edge states with neutral fermions in synthetic Hall ribbons}},\ }\href {https://doi.org/10.1126/science.aaa8736} {\bibfield  {journal} {\bibinfo  {journal} {Science}\ }\textbf {\bibinfo {volume} {349}},\ \bibinfo {pages} {1510} (\bibinfo {year}
  {2015})}\BibitemShut {NoStop}%
\bibitem [{\citenamefont {Atala}\ \emph {et~al.}(2014)\citenamefont {Atala}, \citenamefont {Aidelsburger}, \citenamefont {Lohse}, \citenamefont {Barreiro}, \citenamefont {Paredes},\ and\ \citenamefont {Bloch}}]{atala2014observation}%
  \BibitemOpen
  \bibfield  {author} {\bibinfo {author} {\bibfnamefont {M.}~\bibnamefont {Atala}}, \bibinfo {author} {\bibfnamefont {M.}~\bibnamefont {Aidelsburger}}, \bibinfo {author} {\bibfnamefont {M.}~\bibnamefont {Lohse}}, \bibinfo {author} {\bibfnamefont {J.~T.}\ \bibnamefont {Barreiro}}, \bibinfo {author} {\bibfnamefont {B.}~\bibnamefont {Paredes}},\ and\ \bibinfo {author} {\bibfnamefont {I.}~\bibnamefont {Bloch}},\ }\bibfield  {title} {\bibinfo {title} {{Observation of chiral currents with ultracold atoms in bosonic ladders}},\ }\href {https://doi.org/10.1038/nphys2998} {\bibfield  {journal} {\bibinfo  {journal} {Nat. Phys.}\ }\textbf {\bibinfo {volume} {10}},\ \bibinfo {pages} {588} (\bibinfo {year} {2014})}\BibitemShut {NoStop}%
\bibitem [{\citenamefont {Impertro}\ \emph {et~al.}(2025)\citenamefont {Impertro}, \citenamefont {Huh}, \citenamefont {Karch}, \citenamefont {Wienand}, \citenamefont {Bloch},\ and\ \citenamefont {Aidelsburger}}]{impertro2024realization}%
  \BibitemOpen
  \bibfield  {author} {\bibinfo {author} {\bibfnamefont {A.}~\bibnamefont {Impertro}}, \bibinfo {author} {\bibfnamefont {S.}~\bibnamefont {Huh}}, \bibinfo {author} {\bibfnamefont {S.}~\bibnamefont {Karch}}, \bibinfo {author} {\bibfnamefont {J.~F.}\ \bibnamefont {Wienand}}, \bibinfo {author} {\bibfnamefont {I.}~\bibnamefont {Bloch}},\ and\ \bibinfo {author} {\bibfnamefont {M.}~\bibnamefont {Aidelsburger}},\ }\bibfield  {title} {\bibinfo {title} {Strongly interacting Meissner phases in large bosonic flux ladders},\ }\href {https://doi.org/10.1038/s41567-025-02890-0} {\bibfield  {journal} {\bibinfo  {journal} {Nature Physics}\ }\textbf {\bibinfo {volume} {21}},\ \bibinfo {pages} {895 } (\bibinfo {year} {2025})}\BibitemShut {NoStop}%
\bibitem [{\citenamefont {Lunt}\ \emph {et~al.}(2024)\citenamefont {Lunt}, \citenamefont {Hill}, \citenamefont {Reiter}, \citenamefont {Preiss}, \citenamefont {Ga{\l}ka},\ and\ \citenamefont {Jochim}}]{lunt2024realization}%
  \BibitemOpen
  \bibfield  {author} {\bibinfo {author} {\bibfnamefont {P.}~\bibnamefont {Lunt}}, \bibinfo {author} {\bibfnamefont {P.}~\bibnamefont {Hill}}, \bibinfo {author} {\bibfnamefont {J.}~\bibnamefont {Reiter}}, \bibinfo {author} {\bibfnamefont {P.~M.}\ \bibnamefont {Preiss}}, \bibinfo {author} {\bibfnamefont {M.}~\bibnamefont {Ga{\l}ka}},\ and\ \bibinfo {author} {\bibfnamefont {S.}~\bibnamefont {Jochim}},\ }\bibfield  {title} {\bibinfo {title} {Realization of a Laughlin state of two rapidly rotating fermions},\ }\href@noop {} {\bibfield  {journal} {\bibinfo  {journal} {Physical Review Letters}\ }\textbf {\bibinfo {volume} {133}},\ \bibinfo {pages} {253401} (\bibinfo {year} {2024})}\BibitemShut {NoStop}%
\bibitem [{\citenamefont {{Kwan}}\ \emph {et~al.}(2026)\citenamefont {{Kwan}}, \citenamefont {{Segura}}, \citenamefont {{Li}}, \citenamefont {{Blatz}}, \citenamefont {{Zhi}}, \citenamefont {{Bakkali-Hassani}}, \citenamefont {{Bohrdt}}, \citenamefont {{Greiter}}, \citenamefont {{Grusdt}},\ and\ \citenamefont {{Greiner}}}]{Joyce2026Pfaffian}%
  \BibitemOpen
  \bibfield  {author} {\bibinfo {author} {\bibfnamefont {J.}~\bibnamefont {{Kwan}}}, \bibinfo {author} {\bibfnamefont {P.}~\bibnamefont {{Segura}}}, \bibinfo {author} {\bibfnamefont {Y.}~\bibnamefont {{Li}}}, \bibinfo {author} {\bibfnamefont {T.}~\bibnamefont {{Blatz}}}, \bibinfo {author} {\bibfnamefont {A.}~\bibnamefont {{Zhi}}}, \bibinfo {author} {\bibfnamefont {B.}~\bibnamefont {{Bakkali-Hassani}}}, \bibinfo {author} {\bibfnamefont {A.}~\bibnamefont {{Bohrdt}}}, \bibinfo {author} {\bibfnamefont {M.}~\bibnamefont {{Greiter}}}, \bibinfo {author} {\bibfnamefont {F.}~\bibnamefont {{Grusdt}}},\ and\ \bibinfo {author} {\bibfnamefont {M.}~\bibnamefont {{Greiner}}},\ }\bibinfo {title} {{A Pfaffian quantum Hall state of ultracold bosons}},\ \Eprint {https://arxiv.org/abs/2606.12409} {arXiv:2606.12409} \BibitemShut {NoStop}%
\bibitem [{\citenamefont {Palm}\ \emph {et~al.}(2021)\citenamefont {Palm}, \citenamefont {Buser}, \citenamefont {L\'eonard}, \citenamefont {Aidelsburger}, \citenamefont {Schollw\"ock},\ and\ \citenamefont {Grusdt}}]{palm2021pfaffian}%
  \BibitemOpen
  \bibfield  {author} {\bibinfo {author} {\bibfnamefont {F.~A.}\ \bibnamefont {Palm}}, \bibinfo {author} {\bibfnamefont {M.}~\bibnamefont {Buser}}, \bibinfo {author} {\bibfnamefont {J.}~\bibnamefont {L\'eonard}}, \bibinfo {author} {\bibfnamefont {M.}~\bibnamefont {Aidelsburger}}, \bibinfo {author} {\bibfnamefont {U.}~\bibnamefont {Schollw\"ock}},\ and\ \bibinfo {author} {\bibfnamefont {F.}~\bibnamefont {Grusdt}},\ }\bibfield  {title} {\bibinfo {title} {{Bosonic Pfaffian state in the Hofstadter-Bose-Hubbard model}},\ }\href {https://doi.org/10.1103/PhysRevB.103.L161101} {\bibfield  {journal} {\bibinfo  {journal} {Phys. Rev. B}\ }\textbf {\bibinfo {volume} {103}},\ \bibinfo {pages} {L161101} (\bibinfo {year} {2021})}\BibitemShut {NoStop}%
\bibitem [{\citenamefont {Boesl}\ \emph {et~al.}(2022)\citenamefont {Boesl}, \citenamefont {Dilip}, \citenamefont {Pollmann},\ and\ \citenamefont {Knap}}]{Boesl_prb2022_topomott}%
  \BibitemOpen
  \bibfield  {author} {\bibinfo {author} {\bibfnamefont {J.}~\bibnamefont {Boesl}}, \bibinfo {author} {\bibfnamefont {R.}~\bibnamefont {Dilip}}, \bibinfo {author} {\bibfnamefont {F.}~\bibnamefont {Pollmann}},\ and\ \bibinfo {author} {\bibfnamefont {M.}~\bibnamefont {Knap}},\ }\bibfield  {title} {\bibinfo {title} {Characterizing fractional topological phases of lattice bosons near the first Mott lobe},\ }\href {https://doi.org/10.1103/PhysRevB.105.075135} {\bibfield  {journal} {\bibinfo  {journal} {Phys. Rev. B}\ }\textbf {\bibinfo {volume} {105}},\ \bibinfo {pages} {075135} (\bibinfo {year} {2022})}\BibitemShut {NoStop}%
\bibitem [{\citenamefont {Binanti}\ \emph {et~al.}(2024)\citenamefont {Binanti}, \citenamefont {Goldman},\ and\ \citenamefont {Repellin}}]{binatiRepellin_prr2024_spectroscopyFQH}%
  \BibitemOpen
  \bibfield  {author} {\bibinfo {author} {\bibfnamefont {F.}~\bibnamefont {Binanti}}, \bibinfo {author} {\bibfnamefont {N.}~\bibnamefont {Goldman}},\ and\ \bibinfo {author} {\bibfnamefont {C.}~\bibnamefont {Repellin}},\ }\bibfield  {title} {\bibinfo {title} {Spectroscopy of edge and bulk collective modes in fractional Chern insulators},\ }\href {https://doi.org/10.1103/PhysRevResearch.6.L012054} {\bibfield  {journal} {\bibinfo  {journal} {Phys. Rev. Res.}\ }\textbf {\bibinfo {volume} {6}},\ \bibinfo {pages} {L012054} (\bibinfo {year} {2024})}\BibitemShut {NoStop}%
\bibitem [{\citenamefont {Nardin}\ \emph {et~al.}(2026)\citenamefont {Nardin}, \citenamefont {Mera}, \citenamefont {Defossez}, \citenamefont {Bermond}, \citenamefont {Ozawa},\ and\ \citenamefont {Goldman}}]{nardin2026hallviscositymetricsensitivedichroic}%
  \BibitemOpen
  \bibfield  {author} {\bibinfo {author} {\bibfnamefont {A.}~\bibnamefont {Nardin}}, \bibinfo {author} {\bibfnamefont {B.}~\bibnamefont {Mera}}, \bibinfo {author} {\bibfnamefont {A.}~\bibnamefont {Defossez}}, \bibinfo {author} {\bibfnamefont {B.}~\bibnamefont {Bermond}}, \bibinfo {author} {\bibfnamefont {T.}~\bibnamefont {Ozawa}},\ and\ \bibinfo {author} {\bibfnamefont {N.}~\bibnamefont {Goldman}},\ }\bibinfo {title} {Hall viscosity from metric-sensitive dichroic probes},\ \Eprint {https://arxiv.org/abs/2606.30051} {arXiv:2606.30051} \BibitemShut {NoStop}%
\bibitem [{\citenamefont {Nardin}\ \emph {et~al.}(2024)\citenamefont {Nardin}, \citenamefont {De~Bernardis}, \citenamefont {Umucal{\ifmmode\imath\else\i\fi}lar}, \citenamefont {Mazza}, \citenamefont {Rizzi},\ and\ \citenamefont {Carusotto}}]{nardin2024quantum}%
  \BibitemOpen
  \bibfield  {author} {\bibinfo {author} {\bibfnamefont {A.}~\bibnamefont {Nardin}}, \bibinfo {author} {\bibfnamefont {D.}~\bibnamefont {De~Bernardis}}, \bibinfo {author} {\bibfnamefont {R.~O.}\ \bibnamefont {Umucal{\ifmmode\imath\else\i\fi}lar}}, \bibinfo {author} {\bibfnamefont {L.}~\bibnamefont {Mazza}}, \bibinfo {author} {\bibfnamefont {M.}~\bibnamefont {Rizzi}},\ and\ \bibinfo {author} {\bibfnamefont {I.}~\bibnamefont {Carusotto}},\ }\bibfield  {title} {\bibinfo {title} {{Quantum Nonlinear Optics on the Edge of a Few-Particle Fractional Quantum Hall Fluid in a Small Lattice}},\ }\href {https://doi.org/10.1103/PhysRevLett.133.183401} {\bibfield  {journal} {\bibinfo  {journal} {Phys. Rev. Lett.}\ }\textbf {\bibinfo {volume} {133}},\ \bibinfo {pages} {183401} (\bibinfo {year} {2024})}\BibitemShut {NoStop}%
\bibitem [{\citenamefont {Gromov}\ and\ \citenamefont {Son}(2017)}]{gromov2017bimetric}%
  \BibitemOpen
  \bibfield  {author} {\bibinfo {author} {\bibfnamefont {A.}~\bibnamefont {Gromov}}\ and\ \bibinfo {author} {\bibfnamefont {D.~T.}\ \bibnamefont {Son}},\ }\bibfield  {title} {\bibinfo {title} {{Bimetric Theory of Fractional Quantum Hall States}},\ }\href {https://doi.org/10.1103/PhysRevX.7.041032} {\bibfield  {journal} {\bibinfo  {journal} {Phys. Rev. X}\ }\textbf {\bibinfo {volume} {7}},\ \bibinfo {pages} {041032} (\bibinfo {year} {2017})}\BibitemShut {NoStop}%
\bibitem [{\citenamefont {Liu}\ \emph {et~al.}(2018)\citenamefont {Liu}, \citenamefont {Gromov},\ and\ \citenamefont {Papi\ifmmode~\acute{c}\else \'{c}\fi{}}}]{liu2018quench}%
  \BibitemOpen
  \bibfield  {author} {\bibinfo {author} {\bibfnamefont {Z.}~\bibnamefont {Liu}}, \bibinfo {author} {\bibfnamefont {A.}~\bibnamefont {Gromov}},\ and\ \bibinfo {author} {\bibfnamefont {Z.}~\bibnamefont {Papi\ifmmode~\acute{c}\else \'{c}\fi{}}},\ }\bibfield  {title} {\bibinfo {title} {Geometric quench and nonequilibrium dynamics of fractional quantum Hall states},\ }\href {https://doi.org/10.1103/PhysRevB.98.155140} {\bibfield  {journal} {\bibinfo  {journal} {Phys. Rev. B}\ }\textbf {\bibinfo {volume} {98}},\ \bibinfo {pages} {155140} (\bibinfo {year} {2018})}\BibitemShut {NoStop}%
\bibitem [{\citenamefont {Liou}\ \emph {et~al.}(2019)\citenamefont {Liou}, \citenamefont {Haldane}, \citenamefont {Yang},\ and\ \citenamefont {Rezayi}}]{liou2019chiral}%
  \BibitemOpen
  \bibfield  {author} {\bibinfo {author} {\bibfnamefont {S.-F.}\ \bibnamefont {Liou}}, \bibinfo {author} {\bibfnamefont {F.~D.~M.}\ \bibnamefont {Haldane}}, \bibinfo {author} {\bibfnamefont {K.}~\bibnamefont {Yang}},\ and\ \bibinfo {author} {\bibfnamefont {E.~H.}\ \bibnamefont {Rezayi}},\ }\bibfield  {title} {\bibinfo {title} {{Chiral Gravitons in Fractional Quantum Hall Liquids}},\ }\href {https://doi.org/10.1103/PhysRevLett.123.146801} {\bibfield  {journal} {\bibinfo  {journal} {Phys. Rev. Lett.}\ }\textbf {\bibinfo {volume} {123}},\ \bibinfo {pages} {146801} (\bibinfo {year} {2019})}\BibitemShut {NoStop}%
\bibitem [{\citenamefont {Nguyen}\ \emph {et~al.}(2022)\citenamefont {Nguyen}, \citenamefont {Haldane}, \citenamefont {Rezayi}, \citenamefont {Son},\ and\ \citenamefont {Yang}}]{nguyen2022multiple}%
  \BibitemOpen
  \bibfield  {author} {\bibinfo {author} {\bibfnamefont {D.~X.}\ \bibnamefont {Nguyen}}, \bibinfo {author} {\bibfnamefont {F.~D.~M.}\ \bibnamefont {Haldane}}, \bibinfo {author} {\bibfnamefont {E.~H.}\ \bibnamefont {Rezayi}}, \bibinfo {author} {\bibfnamefont {D.~T.}\ \bibnamefont {Son}},\ and\ \bibinfo {author} {\bibfnamefont {K.}~\bibnamefont {Yang}},\ }\bibfield  {title} {\bibinfo {title} {{Multiple Magnetorotons and Spectral Sum Rules in Fractional Quantum Hall Systems}},\ }\href {https://doi.org/10.1103/PhysRevLett.128.246402} {\bibfield  {journal} {\bibinfo  {journal} {Phys. Rev. Lett.}\ }\textbf {\bibinfo {volume} {128}},\ \bibinfo {pages} {246402} (\bibinfo {year} {2022})}\BibitemShut {NoStop}%
\bibitem [{\citenamefont {Kumar}\ and\ \citenamefont {Haldane}(2022)}]{kumar2022neutral}%
  \BibitemOpen
  \bibfield  {author} {\bibinfo {author} {\bibfnamefont {P.}~\bibnamefont {Kumar}}\ and\ \bibinfo {author} {\bibfnamefont {F.~D.~M.}\ \bibnamefont {Haldane}},\ }\bibfield  {title} {\bibinfo {title} {Neutral excitations of quantum Hall states: A density matrix renormalization group study},\ }\href {https://doi.org/10.1103/PhysRevB.106.075116} {\bibfield  {journal} {\bibinfo  {journal} {Phys. Rev. B}\ }\textbf {\bibinfo {volume} {106}},\ \bibinfo {pages} {075116} (\bibinfo {year} {2022})}\BibitemShut {NoStop}%
\bibitem [{\citenamefont {Yang}(2016)}]{yang2016acoustic}%
  \BibitemOpen
  \bibfield  {author} {\bibinfo {author} {\bibfnamefont {K.}~\bibnamefont {Yang}},\ }\bibfield  {title} {\bibinfo {title} {{Acoustic wave absorption as a probe of dynamical geometrical response of fractional quantum Hall liquids}},\ }\href {https://doi.org/10.1103/PhysRevB.93.161302} {\bibfield  {journal} {\bibinfo  {journal} {Phys. Rev. B}\ }\textbf {\bibinfo {volume} {93}},\ \bibinfo {pages} {161302} (\bibinfo {year} {2016})}\BibitemShut {NoStop}%
\bibitem [{\citenamefont {Liu}\ \emph {et~al.}(2024{\natexlab{a}})\citenamefont {Liu}, \citenamefont {Zhao},\ and\ \citenamefont {Xiang}}]{liuResolving2024}%
  \BibitemOpen
  \bibfield  {author} {\bibinfo {author} {\bibfnamefont {Y.}~\bibnamefont {Liu}}, \bibinfo {author} {\bibfnamefont {T.}~\bibnamefont {Zhao}},\ and\ \bibinfo {author} {\bibfnamefont {T.}~\bibnamefont {Xiang}},\ }\bibfield  {title} {\bibinfo {title} {Resolving geometric excitations of fractional quantum Hall states},\ }\href {https://doi.org/10.1103/PhysRevB.110.195137} {\bibfield  {journal} {\bibinfo  {journal} {Phys. Rev. B}\ }\textbf {\bibinfo {volume} {110}},\ \bibinfo {pages} {195137} (\bibinfo {year} {2024}{\natexlab{a}})}\BibitemShut {NoStop}%
\bibitem [{\citenamefont {Balram}\ \emph {et~al.}(2024)\citenamefont {Balram}, \citenamefont {Sreejith},\ and\ \citenamefont {Jain}}]{Balram_2024}%
  \BibitemOpen
  \bibfield  {author} {\bibinfo {author} {\bibfnamefont {A.~C.}\ \bibnamefont {Balram}}, \bibinfo {author} {\bibfnamefont {G.}~\bibnamefont {Sreejith}},\ and\ \bibinfo {author} {\bibfnamefont {J.}~\bibnamefont {Jain}},\ }\bibfield  {title} {\bibinfo {title} {Splitting of the Girvin-MacDonald-Platzman Density Wave and the Nature of Chiral Gravitons in the Fractional Quantum Hall Effect},\ }\bibfield  {journal} {\bibinfo  {journal} {Physical Review Letters}\ }\textbf {\bibinfo {volume} {133}},\ \href {https://doi.org/10.1103/physrevlett.133.246605} {10.1103/physrevlett.133.246605} (\bibinfo {year} {2024})\BibitemShut {NoStop}%
\bibitem [{\citenamefont {Yang}\ \emph {et~al.}(2012)\citenamefont {Yang}, \citenamefont {Hu}, \citenamefont {Papi\ifmmode~\acute{c}\else \'{c}\fi{}},\ and\ \citenamefont {Haldane}}]{Yang_prl2012_modelwavefunctions_graviton}%
  \BibitemOpen
  \bibfield  {author} {\bibinfo {author} {\bibfnamefont {B.}~\bibnamefont {Yang}}, \bibinfo {author} {\bibfnamefont {Z.-X.}\ \bibnamefont {Hu}}, \bibinfo {author} {\bibfnamefont {Z.}~\bibnamefont {Papi\ifmmode~\acute{c}\else \'{c}\fi{}}},\ and\ \bibinfo {author} {\bibfnamefont {F.~D.~M.}\ \bibnamefont {Haldane}},\ }\bibfield  {title} {\bibinfo {title} {Model Wave Functions for the Collective Modes and the Magnetoroton Theory of the Fractional Quantum Hall Effect},\ }\href {https://doi.org/10.1103/PhysRevLett.108.256807} {\bibfield  {journal} {\bibinfo  {journal} {Phys. Rev. Lett.}\ }\textbf {\bibinfo {volume} {108}},\ \bibinfo {pages} {256807} (\bibinfo {year} {2012})}\BibitemShut {NoStop}%
\bibitem [{\citenamefont {Nguyen}\ and\ \citenamefont {Son}(2021)}]{Nguyen2021Dirac}%
  \BibitemOpen
  \bibfield  {author} {\bibinfo {author} {\bibfnamefont {D.~X.}\ \bibnamefont {Nguyen}}\ and\ \bibinfo {author} {\bibfnamefont {D.~T.}\ \bibnamefont {Son}},\ }\bibfield  {title} {\bibinfo {title} {Dirac composite fermion theory of general Jain sequences},\ }\bibfield  {journal} {\bibinfo  {journal} {Physical Review Research}\ }\textbf {\bibinfo {volume} {3}},\ \href {https://doi.org/10.1103/physrevresearch.3.033217} {10.1103/physrevresearch.3.033217} (\bibinfo {year} {2021})\BibitemShut {NoStop}%
\bibitem [{\citenamefont {Wang}\ and\ \citenamefont {Yang}(2022)}]{Wang2022Analytic}%
  \BibitemOpen
  \bibfield  {author} {\bibinfo {author} {\bibfnamefont {Y.}~\bibnamefont {Wang}}\ and\ \bibinfo {author} {\bibfnamefont {B.}~\bibnamefont {Yang}},\ }\bibfield  {title} {\bibinfo {title} {Analytic exposition of the graviton modes in fractional quantum Hall effects and its physical implications},\ }\bibfield  {journal} {\bibinfo  {journal} {Physical Review B}\ }\textbf {\bibinfo {volume} {105}},\ \href {https://doi.org/10.1103/physrevb.105.035144} {10.1103/physrevb.105.035144} (\bibinfo {year} {2022})\BibitemShut {NoStop}%
\bibitem [{\citenamefont {Yuzhu}\ and\ \citenamefont {Bo}(2023)}]{wang2023Geometric}%
  \BibitemOpen
  \bibfield  {author} {\bibinfo {author} {\bibfnamefont {W.}~\bibnamefont {Yuzhu}}\ and\ \bibinfo {author} {\bibfnamefont {Y.}~\bibnamefont {Bo}},\ }\bibfield  {title} {\bibinfo {title} {Geometric fluctuation of conformal Hilbert spaces and multiple graviton modes in fractional quantum Hall effect},\ }\bibfield  {journal} {\bibinfo  {journal} {Nature Communications}\ }\textbf {\bibinfo {volume} {14}},\ \href {https://doi.org/10.1038/s41467-023-38036-0} {10.1038/s41467-023-38036-0} (\bibinfo {year} {2023})\BibitemShut {NoStop}%
\bibitem [{\citenamefont {Haldane}\ \emph {et~al.}(2021)\citenamefont {Haldane}, \citenamefont {Rezayi},\ and\ \citenamefont {Yang}}]{Haldane2021Graviton}%
  \BibitemOpen
  \bibfield  {author} {\bibinfo {author} {\bibfnamefont {F.~D.~M.}\ \bibnamefont {Haldane}}, \bibinfo {author} {\bibfnamefont {E.~H.}\ \bibnamefont {Rezayi}},\ and\ \bibinfo {author} {\bibfnamefont {K.}~\bibnamefont {Yang}},\ }\bibfield  {title} {\bibinfo {title} {Graviton chirality and topological order in the half-filled Landau level},\ }\href {https://doi.org/10.1103/PhysRevB.104.L121106} {\bibfield  {journal} {\bibinfo  {journal} {Phys. Rev. B}\ }\textbf {\bibinfo {volume} {104}},\ \bibinfo {pages} {L121106} (\bibinfo {year} {2021})}\BibitemShut {NoStop}%
\bibitem [{\citenamefont {Pu}\ \emph {et~al.}(2024)\citenamefont {Pu}, \citenamefont {Balram}, \citenamefont {Taylor}, \citenamefont {Fradkin},\ and\ \citenamefont {Papi\ifmmode~\acute{c}\else \'{c}\fi{}}}]{pu2024microscopic}%
  \BibitemOpen
  \bibfield  {author} {\bibinfo {author} {\bibfnamefont {S.}~\bibnamefont {Pu}}, \bibinfo {author} {\bibfnamefont {A.~C.}\ \bibnamefont {Balram}}, \bibinfo {author} {\bibfnamefont {J.}~\bibnamefont {Taylor}}, \bibinfo {author} {\bibfnamefont {E.}~\bibnamefont {Fradkin}},\ and\ \bibinfo {author} {\bibfnamefont {Z.}~\bibnamefont {Papi\ifmmode~\acute{c}\else \'{c}\fi{}}},\ }\bibfield  {title} {\bibinfo {title} {{Microscopic Model for Fractional Quantum Hall Nematics}},\ }\href {https://doi.org/10.1103/PhysRevLett.132.236503} {\bibfield  {journal} {\bibinfo  {journal} {Phys. Rev. Lett.}\ }\textbf {\bibinfo {volume} {132}},\ \bibinfo {pages} {236503} (\bibinfo {year} {2024})}\BibitemShut {NoStop}%
\bibitem [{\citenamefont {Bacciconi}\ \emph {et~al.}(2025)\citenamefont {Bacciconi}, \citenamefont {Xavier}, \citenamefont {Carusotto}, \citenamefont {Chanda},\ and\ \citenamefont {Dalmonte}}]{bacciconi_prx2025_gravitonpolaritons}%
  \BibitemOpen
  \bibfield  {author} {\bibinfo {author} {\bibfnamefont {Z.}~\bibnamefont {Bacciconi}}, \bibinfo {author} {\bibfnamefont {H.~B.}\ \bibnamefont {Xavier}}, \bibinfo {author} {\bibfnamefont {I.}~\bibnamefont {Carusotto}}, \bibinfo {author} {\bibfnamefont {T.}~\bibnamefont {Chanda}},\ and\ \bibinfo {author} {\bibfnamefont {M.}~\bibnamefont {Dalmonte}},\ }\bibfield  {title} {\bibinfo {title} {Theory of Fractional Quantum Hall Liquids Coupled to Quantum Light and Emergent Graviton-Polaritons},\ }\href {https://doi.org/10.1103/PhysRevX.15.021027} {\bibfield  {journal} {\bibinfo  {journal} {Phys. Rev. X}\ }\textbf {\bibinfo {volume} {15}},\ \bibinfo {pages} {021027} (\bibinfo {year} {2025})}\BibitemShut {NoStop}%
\bibitem [{\citenamefont {Balram}\ \emph {et~al.}(2022)\citenamefont {Balram}, \citenamefont {Liu}, \citenamefont {Gromov},\ and\ \citenamefont {Papi\ifmmode~\acute{c}\else \'{c}\fi{}}}]{AjitPapic_prx2022_highpartons}%
  \BibitemOpen
  \bibfield  {author} {\bibinfo {author} {\bibfnamefont {A.~C.}\ \bibnamefont {Balram}}, \bibinfo {author} {\bibfnamefont {Z.}~\bibnamefont {Liu}}, \bibinfo {author} {\bibfnamefont {A.}~\bibnamefont {Gromov}},\ and\ \bibinfo {author} {\bibfnamefont {Z.}~\bibnamefont {Papi\ifmmode~\acute{c}\else \'{c}\fi{}}},\ }\bibfield  {title} {\bibinfo {title} {Very-High-Energy Collective States of Partons in Fractional Quantum Hall Liquids},\ }\href {https://doi.org/10.1103/PhysRevX.12.021008} {\bibfield  {journal} {\bibinfo  {journal} {Phys. Rev. X}\ }\textbf {\bibinfo {volume} {12}},\ \bibinfo {pages} {021008} (\bibinfo {year} {2022})}\BibitemShut {NoStop}%
\bibitem [{\citenamefont {Girvin}\ \emph {et~al.}(1986)\citenamefont {Girvin}, \citenamefont {MacDonald},\ and\ \citenamefont {Platzman}}]{gmp1986magnetoroton}%
  \BibitemOpen
  \bibfield  {author} {\bibinfo {author} {\bibfnamefont {S.~M.}\ \bibnamefont {Girvin}}, \bibinfo {author} {\bibfnamefont {A.~H.}\ \bibnamefont {MacDonald}},\ and\ \bibinfo {author} {\bibfnamefont {P.~M.}\ \bibnamefont {Platzman}},\ }\bibfield  {title} {\bibinfo {title} {{Magneto-roton theory of collective excitations in the fractional quantum Hall effect}},\ }\href {https://doi.org/10.1103/PhysRevB.33.2481} {\bibfield  {journal} {\bibinfo  {journal} {Phys. Rev. B}\ }\textbf {\bibinfo {volume} {33}},\ \bibinfo {pages} {2481} (\bibinfo {year} {1986})}\BibitemShut {NoStop}%
\bibitem [{\citenamefont {M\"{o}ller}\ \emph {et~al.}(2011)\citenamefont {M\"{o}ller}, \citenamefont {Wójs},\ and\ \citenamefont {Cooper}}]{Moller2011neutral}%
  \BibitemOpen
  \bibfield  {author} {\bibinfo {author} {\bibfnamefont {G.}~\bibnamefont {M\"{o}ller}}, \bibinfo {author} {\bibfnamefont {A.}~\bibnamefont {Wójs}},\ and\ \bibinfo {author} {\bibfnamefont {N.~R.}\ \bibnamefont {Cooper}},\ }\bibfield  {title} {\bibinfo {title} {Neutral Fermion Excitations in the Moore-Read State at Filling Factor $\nu=5/2$},\ }\bibfield  {journal} {\bibinfo  {journal} {Physical Review Letters}\ }\textbf {\bibinfo {volume} {107}},\ \href {https://doi.org/10.1103/physrevlett.107.036803} {10.1103/physrevlett.107.036803} (\bibinfo {year} {2011})\BibitemShut {NoStop}%
\bibitem [{\citenamefont {Bonderson}\ \emph {et~al.}(2011)\citenamefont {Bonderson}, \citenamefont {Feiguin},\ and\ \citenamefont {Nayak}}]{Bonderson2011Numerical}%
  \BibitemOpen
  \bibfield  {author} {\bibinfo {author} {\bibfnamefont {P.}~\bibnamefont {Bonderson}}, \bibinfo {author} {\bibfnamefont {A.~E.}\ \bibnamefont {Feiguin}},\ and\ \bibinfo {author} {\bibfnamefont {C.}~\bibnamefont {Nayak}},\ }\bibfield  {title} {\bibinfo {title} {Numerical Calculation of the Neutral Fermion Gap at the $\nu=5/2$ Fractional Quantum Hall State},\ }\bibfield  {journal} {\bibinfo  {journal} {Physical Review Letters}\ }\textbf {\bibinfo {volume} {106}},\ \href {https://doi.org/10.1103/physrevlett.106.186802} {10.1103/physrevlett.106.186802} (\bibinfo {year} {2011})\BibitemShut {NoStop}%
\bibitem [{\citenamefont {Repellin}\ \emph {et~al.}(2015)\citenamefont {Repellin}, \citenamefont {Neupert}, \citenamefont {Bernevig},\ and\ \citenamefont {Regnault}}]{Repellin2015projective}%
  \BibitemOpen
  \bibfield  {author} {\bibinfo {author} {\bibfnamefont {C.}~\bibnamefont {Repellin}}, \bibinfo {author} {\bibfnamefont {T.}~\bibnamefont {Neupert}}, \bibinfo {author} {\bibfnamefont {B.~A.}\ \bibnamefont {Bernevig}},\ and\ \bibinfo {author} {\bibfnamefont {N.}~\bibnamefont {Regnault}},\ }\bibfield  {title} {\bibinfo {title} {Projective construction of the ${\mathbb{Z}}_{k}$ Read-Rezayi fractional quantum Hall states and their excitations on the torus geometry},\ }\href {https://doi.org/10.1103/PhysRevB.92.115128} {\bibfield  {journal} {\bibinfo  {journal} {Phys. Rev. B}\ }\textbf {\bibinfo {volume} {92}},\ \bibinfo {pages} {115128} (\bibinfo {year} {2015})}\BibitemShut {NoStop}%
\bibitem [{\citenamefont {{Long}}\ \emph {et~al.}(2026)\citenamefont {{Long}}, \citenamefont {{Bacciconi}}, \citenamefont {{Lu}}, \citenamefont {{Xavier}}, \citenamefont {{Meng}},\ and\ \citenamefont {{Dalmonte}}}]{longChiral2026}%
  \BibitemOpen
  \bibfield  {author} {\bibinfo {author} {\bibfnamefont {M.}~\bibnamefont {{Long}}}, \bibinfo {author} {\bibfnamefont {Z.}~\bibnamefont {{Bacciconi}}}, \bibinfo {author} {\bibfnamefont {H.}~\bibnamefont {{Lu}}}, \bibinfo {author} {\bibfnamefont {H.~B.}\ \bibnamefont {{Xavier}}}, \bibinfo {author} {\bibfnamefont {Z.~Y.}\ \bibnamefont {{Meng}}},\ and\ \bibinfo {author} {\bibfnamefont {M.}~\bibnamefont {{Dalmonte}}},\ }\bibfield  {title} {\bibinfo {title} {{Chiral Graviton Modes in Fermionic Fractional Chern Insulators}},\ }\href {https://doi.org/10.48550/arXiv.2601.05196} {\bibfield  {journal} {\bibinfo  {journal} {arXiv e-prints}\ ,\ \bibinfo {eid} {arXiv:2601.05196}} (\bibinfo {year} {2026})}\BibitemShut {NoStop}%
\bibitem [{\citenamefont {Daley}\ \emph {et~al.}(2009)\citenamefont {Daley}, \citenamefont {Taylor}, \citenamefont {Diehl}, \citenamefont {Baranov},\ and\ \citenamefont {Zoller}}]{daley2009atomic}%
  \BibitemOpen
  \bibfield  {author} {\bibinfo {author} {\bibfnamefont {A.~J.}\ \bibnamefont {Daley}}, \bibinfo {author} {\bibfnamefont {J.~M.}\ \bibnamefont {Taylor}}, \bibinfo {author} {\bibfnamefont {S.}~\bibnamefont {Diehl}}, \bibinfo {author} {\bibfnamefont {M.}~\bibnamefont {Baranov}},\ and\ \bibinfo {author} {\bibfnamefont {P.}~\bibnamefont {Zoller}},\ }\bibfield  {title} {\bibinfo {title} {Atomic Three-Body Loss as a Dynamical Three-Body Interaction},\ }\href {https://doi.org/10.1103/PhysRevLett.102.040402} {\bibfield  {journal} {\bibinfo  {journal} {Phys. Rev. Lett.}\ }\textbf {\bibinfo {volume} {102}},\ \bibinfo {pages} {040402} (\bibinfo {year} {2009})}\BibitemShut {NoStop}%
\bibitem [{\citenamefont {Regnault}\ and\ \citenamefont {Jolicoeur}(2003)}]{regnault_prl2003_bosonsV0}%
  \BibitemOpen
  \bibfield  {author} {\bibinfo {author} {\bibfnamefont {N.}~\bibnamefont {Regnault}}\ and\ \bibinfo {author} {\bibfnamefont {T.}~\bibnamefont {Jolicoeur}},\ }\bibfield  {title} {\bibinfo {title} {Quantum Hall Fractions in Rotating Bose-Einstein Condensates},\ }\href {https://doi.org/10.1103/PhysRevLett.91.030402} {\bibfield  {journal} {\bibinfo  {journal} {Phys. Rev. Lett.}\ }\textbf {\bibinfo {volume} {91}},\ \bibinfo {pages} {030402} (\bibinfo {year} {2003})}\BibitemShut {NoStop}%
\bibitem [{\citenamefont {Long}\ \emph {et~al.}(2026)\citenamefont {Long}, \citenamefont {Lu}, \citenamefont {Wu},\ and\ \citenamefont {Meng}}]{longSpectra2025}%
  \BibitemOpen
  \bibfield  {author} {\bibinfo {author} {\bibfnamefont {M.}~\bibnamefont {Long}}, \bibinfo {author} {\bibfnamefont {H.}~\bibnamefont {Lu}}, \bibinfo {author} {\bibfnamefont {H.-Q.}\ \bibnamefont {Wu}},\ and\ \bibinfo {author} {\bibfnamefont {Z.~Y.}\ \bibnamefont {Meng}},\ }\bibfield  {title} {\bibinfo {title} {Spectra of magnetoroton and chiral graviton modes of the fractional Chern insulator},\ }\bibfield  {journal} {\bibinfo  {journal} {Physical Review B}\ }\textbf {\bibinfo {volume} {113}},\ \href {https://doi.org/Phys. Rev. B 113, L041108} {Phys. Rev. B 113, L041108} (\bibinfo {year} {2026})\BibitemShut {NoStop}%
\bibitem [{sup()}]{suppmat}%
  \BibitemOpen
  \bibinfo {note} {See Supplemental Material for extra ED and MPS results, including results for fluxes $n_\phi=1/7$ and $1/5$, weights for smaller droplets, and a brief phase diagram characterization on cylinders.}\BibitemShut {Stop}%
\bibitem [{\citenamefont {White}(1992)}]{White1992}%
  \BibitemOpen
  \bibfield  {author} {\bibinfo {author} {\bibfnamefont {S.~R.}\ \bibnamefont {White}},\ }\bibfield  {title} {\bibinfo {title} {{Density matrix formulation for quantum renormalization groups}},\ }\href {https://doi.org/10.1103/PhysRevLett.69.2863} {\bibfield  {journal} {\bibinfo  {journal} {Phys. Rev. Lett.}\ }\textbf {\bibinfo {volume} {69}},\ \bibinfo {pages} {2863} (\bibinfo {year} {1992})}\BibitemShut {NoStop}%
\bibitem [{\citenamefont {White}(1993)}]{White1993}%
  \BibitemOpen
  \bibfield  {author} {\bibinfo {author} {\bibfnamefont {S.~R.}\ \bibnamefont {White}},\ }\bibfield  {title} {\bibinfo {title} {{Density-matrix algorithms for quantum renormalization groups}},\ }\href {https://doi.org/10.1103/PhysRevB.48.10345} {\bibfield  {journal} {\bibinfo  {journal} {Phys. Rev. B}\ }\textbf {\bibinfo {volume} {48}},\ \bibinfo {pages} {10345} (\bibinfo {year} {1993})}\BibitemShut {NoStop}%
\bibitem [{\citenamefont {Schollw\"{o}ck}(2011)}]{Schollwock2011}%
  \BibitemOpen
  \bibfield  {author} {\bibinfo {author} {\bibfnamefont {U.}~\bibnamefont {Schollw\"{o}ck}},\ }\bibfield  {title} {\bibinfo {title} {{The density-matrix renormalization group in the age of matrix product states}},\ }\href {https://doi.org/10.1016/j.aop.2010.09.012} {\bibfield  {journal} {\bibinfo  {journal} {Ann. Phys.}\ }\textbf {\bibinfo {volume} {326}},\ \bibinfo {pages} {96} (\bibinfo {year} {2011})}\BibitemShut {NoStop}%
\bibitem [{\citenamefont {Haegeman}\ \emph {et~al.}(2011{\natexlab{a}})\citenamefont {Haegeman}, \citenamefont {Cirac}, \citenamefont {Osborne}, \citenamefont {Pi\ifmmode~\check{z}\else \v{z}\fi{}orn}, \citenamefont {Verschelde},\ and\ \citenamefont {Verstraete}}]{Haegeman2011Time}%
  \BibitemOpen
  \bibfield  {author} {\bibinfo {author} {\bibfnamefont {J.}~\bibnamefont {Haegeman}}, \bibinfo {author} {\bibfnamefont {J.~I.}\ \bibnamefont {Cirac}}, \bibinfo {author} {\bibfnamefont {T.~J.}\ \bibnamefont {Osborne}}, \bibinfo {author} {\bibfnamefont {I.}~\bibnamefont {Pi\ifmmode~\check{z}\else \v{z}\fi{}orn}}, \bibinfo {author} {\bibfnamefont {H.}~\bibnamefont {Verschelde}},\ and\ \bibinfo {author} {\bibfnamefont {F.}~\bibnamefont {Verstraete}},\ }\bibfield  {title} {\bibinfo {title} {Time-Dependent Variational Principle for Quantum Lattices},\ }\href {https://doi.org/10.1103/PhysRevLett.107.070601} {\bibfield  {journal} {\bibinfo  {journal} {Phys. Rev. Lett.}\ }\textbf {\bibinfo {volume} {107}},\ \bibinfo {pages} {070601} (\bibinfo {year} {2011}{\natexlab{a}})}\BibitemShut {NoStop}%
\bibitem [{\citenamefont {Haegeman}\ \emph {et~al.}(2016{\natexlab{a}})\citenamefont {Haegeman}, \citenamefont {Lubich}, \citenamefont {Oseledets}, \citenamefont {Vandereycken},\ and\ \citenamefont {Verstraete}}]{Haegeman2016Unifying}%
  \BibitemOpen
  \bibfield  {author} {\bibinfo {author} {\bibfnamefont {J.}~\bibnamefont {Haegeman}}, \bibinfo {author} {\bibfnamefont {C.}~\bibnamefont {Lubich}}, \bibinfo {author} {\bibfnamefont {I.}~\bibnamefont {Oseledets}}, \bibinfo {author} {\bibfnamefont {B.}~\bibnamefont {Vandereycken}},\ and\ \bibinfo {author} {\bibfnamefont {F.}~\bibnamefont {Verstraete}},\ }\bibfield  {title} {\bibinfo {title} {Unifying time evolution and optimization with matrix product states},\ }\href {https://doi.org/10.1103/PhysRevB.94.165116} {\bibfield  {journal} {\bibinfo  {journal} {Phys. Rev. B}\ }\textbf {\bibinfo {volume} {94}},\ \bibinfo {pages} {165116} (\bibinfo {year} {2016}{\natexlab{a}})}\BibitemShut {NoStop}%
\bibitem [{\citenamefont {Xavier}\ \emph {et~al.}(2025)\citenamefont {Xavier}, \citenamefont {Bacciconi}, \citenamefont {Chanda}, \citenamefont {Son},\ and\ \citenamefont {Dalmonte}}]{xavier2025chiralgravitonslattice}%
  \BibitemOpen
  \bibfield  {author} {\bibinfo {author} {\bibfnamefont {H.~B.}\ \bibnamefont {Xavier}}, \bibinfo {author} {\bibfnamefont {Z.}~\bibnamefont {Bacciconi}}, \bibinfo {author} {\bibfnamefont {T.}~\bibnamefont {Chanda}}, \bibinfo {author} {\bibfnamefont {D.~T.}\ \bibnamefont {Son}},\ and\ \bibinfo {author} {\bibfnamefont {M.}~\bibnamefont {Dalmonte}},\ }\bibfield  {title} {\bibinfo {title} {Chiral Graviton Modes on the Lattice},\ }\href {https://doi.org/10.1103/1636-kl65} {\bibfield  {journal} {\bibinfo  {journal} {Phys. Rev. Lett.}\ }\textbf {\bibinfo {volume} {135}},\ \bibinfo {pages} {196501} (\bibinfo {year} {2025})}\BibitemShut {NoStop}%
\bibitem [{\citenamefont {L{\ifmmode\acute{e}\else\'{e}\fi}onard}\ \emph {et~al.}(2023)\citenamefont {L{\ifmmode\acute{e}\else\'{e}\fi}onard}, \citenamefont {Kim}, \citenamefont {Kwan}, \citenamefont {Segura}, \citenamefont {Grusdt}, \citenamefont {Repellin}, \citenamefont {Goldman},\ and\ \citenamefont {Greiner}}]{leonard2023realization}%
  \BibitemOpen
  \bibfield  {author} {\bibinfo {author} {\bibfnamefont {J.}~\bibnamefont {L{\ifmmode\acute{e}\else\'{e}\fi}onard}}, \bibinfo {author} {\bibfnamefont {S.}~\bibnamefont {Kim}}, \bibinfo {author} {\bibfnamefont {J.}~\bibnamefont {Kwan}}, \bibinfo {author} {\bibfnamefont {P.}~\bibnamefont {Segura}}, \bibinfo {author} {\bibfnamefont {F.}~\bibnamefont {Grusdt}}, \bibinfo {author} {\bibfnamefont {C.}~\bibnamefont {Repellin}}, \bibinfo {author} {\bibfnamefont {N.}~\bibnamefont {Goldman}},\ and\ \bibinfo {author} {\bibfnamefont {M.}~\bibnamefont {Greiner}},\ }\bibfield  {title} {\bibinfo {title} {{Realization of a fractional quantum Hall state with ultracold atoms}},\ }\href {https://doi.org/10.1038/s41586-023-06122-4} {\bibfield  {journal} {\bibinfo  {journal} {Nature}\ }\textbf {\bibinfo {volume} {619}},\ \bibinfo {pages} {495} (\bibinfo {year} {2023})}\BibitemShut {NoStop}%
\bibitem [{\citenamefont {Pollmann}\ and\ \citenamefont {Turner}(2012)}]{Pollmann2012detection}%
  \BibitemOpen
  \bibfield  {author} {\bibinfo {author} {\bibfnamefont {F.}~\bibnamefont {Pollmann}}\ and\ \bibinfo {author} {\bibfnamefont {A.~M.}\ \bibnamefont {Turner}},\ }\bibfield  {title} {\bibinfo {title} {Detection of symmetry-protected topological phases in one dimension},\ }\bibfield  {journal} {\bibinfo  {journal} {Physical Review B}\ }\textbf {\bibinfo {volume} {86}},\ \href {https://doi.org/10.1103/physrevb.86.125441} {10.1103/physrevb.86.125441} (\bibinfo {year} {2012})\BibitemShut {NoStop}%
\bibitem [{\citenamefont {Cincio}\ and\ \citenamefont {Vidal}(2013)}]{Cincio2013Characterizing}%
  \BibitemOpen
  \bibfield  {author} {\bibinfo {author} {\bibfnamefont {L.}~\bibnamefont {Cincio}}\ and\ \bibinfo {author} {\bibfnamefont {G.}~\bibnamefont {Vidal}},\ }\bibfield  {title} {\bibinfo {title} {Characterizing Topological Order by Studying the Ground States on an Infinite Cylinder},\ }\bibfield  {journal} {\bibinfo  {journal} {Physical Review Letters}\ }\textbf {\bibinfo {volume} {110}},\ \href {https://doi.org/10.1103/physrevlett.110.067208} {10.1103/physrevlett.110.067208} (\bibinfo {year} {2013})\BibitemShut {NoStop}%
\bibitem [{\citenamefont {Phien}\ \emph {et~al.}(2012)\citenamefont {Phien}, \citenamefont {Vidal},\ and\ \citenamefont {McCulloch}}]{Phien2012infinite}%
  \BibitemOpen
  \bibfield  {author} {\bibinfo {author} {\bibfnamefont {H.~N.}\ \bibnamefont {Phien}}, \bibinfo {author} {\bibfnamefont {G.}~\bibnamefont {Vidal}},\ and\ \bibinfo {author} {\bibfnamefont {I.~P.}\ \bibnamefont {McCulloch}},\ }\bibfield  {title} {\bibinfo {title} {Infinite boundary conditions for matrix product state calculations},\ }\bibfield  {journal} {\bibinfo  {journal} {Physical Review B}\ }\textbf {\bibinfo {volume} {86}},\ \href {https://doi.org/10.1103/physrevb.86.245107} {10.1103/physrevb.86.245107} (\bibinfo {year} {2012})\BibitemShut {NoStop}%
\bibitem [{\citenamefont {Milsted}\ \emph {et~al.}(2013)\citenamefont {Milsted}, \citenamefont {Haegeman}, \citenamefont {Osborne},\ and\ \citenamefont {Verstraete}}]{Milsted2013Variational}%
  \BibitemOpen
  \bibfield  {author} {\bibinfo {author} {\bibfnamefont {A.}~\bibnamefont {Milsted}}, \bibinfo {author} {\bibfnamefont {J.}~\bibnamefont {Haegeman}}, \bibinfo {author} {\bibfnamefont {T.~J.}\ \bibnamefont {Osborne}},\ and\ \bibinfo {author} {\bibfnamefont {F.}~\bibnamefont {Verstraete}},\ }\bibfield  {title} {\bibinfo {title} {Variational matrix product ansatz for nonuniform dynamics in the thermodynamic limit},\ }\bibfield  {journal} {\bibinfo  {journal} {Physical Review B}\ }\textbf {\bibinfo {volume} {88}},\ \href {https://doi.org/10.1103/physrevb.88.155116} {10.1103/physrevb.88.155116} (\bibinfo {year} {2013})\BibitemShut {NoStop}%
\bibitem [{\citenamefont {Liang}\ \emph {et~al.}(2024)\citenamefont {Liang}, \citenamefont {Liu}, \citenamefont {Yang}, \citenamefont {Huang}, \citenamefont {Wurstbauer}, \citenamefont {Dean}, \citenamefont {West}, \citenamefont {Pfeiffer}, \citenamefont {Du},\ and\ \citenamefont {Pinczuk}}]{liang2024evidence}%
  \BibitemOpen
  \bibfield  {author} {\bibinfo {author} {\bibfnamefont {J.}~\bibnamefont {Liang}}, \bibinfo {author} {\bibfnamefont {Z.}~\bibnamefont {Liu}}, \bibinfo {author} {\bibfnamefont {Z.}~\bibnamefont {Yang}}, \bibinfo {author} {\bibfnamefont {Y.}~\bibnamefont {Huang}}, \bibinfo {author} {\bibfnamefont {U.}~\bibnamefont {Wurstbauer}}, \bibinfo {author} {\bibfnamefont {C.~R.}\ \bibnamefont {Dean}}, \bibinfo {author} {\bibfnamefont {K.~W.}\ \bibnamefont {West}}, \bibinfo {author} {\bibfnamefont {L.~N.}\ \bibnamefont {Pfeiffer}}, \bibinfo {author} {\bibfnamefont {L.}~\bibnamefont {Du}},\ and\ \bibinfo {author} {\bibfnamefont {A.}~\bibnamefont {Pinczuk}},\ }\bibfield  {title} {\bibinfo {title} {{Evidence for chiral graviton modes in fractional quantum Hall liquids}},\ }\href {https://doi.org/10.1038/s41586-024-07201-w} {\bibfield  {journal} {\bibinfo  {journal} {Nature}\ }\textbf {\bibinfo {volume} {628}},\ \bibinfo {pages} {78} (\bibinfo {year} {2024})}\BibitemShut {NoStop}%
\bibitem [{\citenamefont {Yang}\ \emph {et~al.}(2026)\citenamefont {Yang}, \citenamefont {Wang}, \citenamefont {Lu}, \citenamefont {Liu}, \citenamefont {Baldwin}, \citenamefont {West}, \citenamefont {Pfeiffer}, \citenamefont {Yang}, \citenamefont {Balram},\ and\ \citenamefont {Du}}]{Yang2026_natphys_gravitonpartons}%
  \BibitemOpen
  \bibfield  {author} {\bibinfo {author} {\bibfnamefont {Z.}~\bibnamefont {Yang}}, \bibinfo {author} {\bibfnamefont {Y.}~\bibnamefont {Wang}}, \bibinfo {author} {\bibfnamefont {X.}~\bibnamefont {Lu}}, \bibinfo {author} {\bibfnamefont {Z.}~\bibnamefont {Liu}}, \bibinfo {author} {\bibfnamefont {K.~W.}\ \bibnamefont {Baldwin}}, \bibinfo {author} {\bibfnamefont {K.~W.}\ \bibnamefont {West}}, \bibinfo {author} {\bibfnamefont {L.~N.}\ \bibnamefont {Pfeiffer}}, \bibinfo {author} {\bibfnamefont {B.}~\bibnamefont {Yang}}, \bibinfo {author} {\bibfnamefont {A.~C.}\ \bibnamefont {Balram}},\ and\ \bibinfo {author} {\bibfnamefont {L.}~\bibnamefont {Du}},\ }\bibfield  {title} {\bibinfo {title} {Emergent partons in fractional quantum Hall systems},\ }\bibfield  {journal} {\bibinfo  {journal} {Nature Physics}\ }\href {https://doi.org/10.1038/s41567-026-03338-9} {10.1038/s41567-026-03338-9} (\bibinfo {year} {2026})\BibitemShut {NoStop}%
\bibitem [{\citenamefont {Liu}\ \emph {et~al.}(2021)\citenamefont {Liu}, \citenamefont {Balram}, \citenamefont {Papi\ifmmode~\acute{c}\else \'{c}\fi{}},\ and\ \citenamefont {Gromov}}]{liu2021quench}%
  \BibitemOpen
  \bibfield  {author} {\bibinfo {author} {\bibfnamefont {Z.}~\bibnamefont {Liu}}, \bibinfo {author} {\bibfnamefont {A.~C.}\ \bibnamefont {Balram}}, \bibinfo {author} {\bibfnamefont {Z.}~\bibnamefont {Papi\ifmmode~\acute{c}\else \'{c}\fi{}}},\ and\ \bibinfo {author} {\bibfnamefont {A.}~\bibnamefont {Gromov}},\ }\bibfield  {title} {\bibinfo {title} {Quench Dynamics of Collective Modes in Fractional Quantum Hall Bilayers},\ }\href {https://doi.org/10.1103/PhysRevLett.126.076604} {\bibfield  {journal} {\bibinfo  {journal} {Phys. Rev. Lett.}\ }\textbf {\bibinfo {volume} {126}},\ \bibinfo {pages} {076604} (\bibinfo {year} {2021})}\BibitemShut {NoStop}%
\bibitem [{\citenamefont {Ippoliti}\ \emph {et~al.}(2018)\citenamefont {Ippoliti}, \citenamefont {Bhatt},\ and\ \citenamefont {Haldane}}]{ippoliti2018geometry}%
  \BibitemOpen
  \bibfield  {author} {\bibinfo {author} {\bibfnamefont {M.}~\bibnamefont {Ippoliti}}, \bibinfo {author} {\bibfnamefont {R.~N.}\ \bibnamefont {Bhatt}},\ and\ \bibinfo {author} {\bibfnamefont {F.~D.~M.}\ \bibnamefont {Haldane}},\ }\bibfield  {title} {\bibinfo {title} {{Geometry of flux attachment in anisotropic fractional quantum Hall states}},\ }\href {https://doi.org/10.1103/PhysRevB.98.085101} {\bibfield  {journal} {\bibinfo  {journal} {Phys. Rev. B}\ }\textbf {\bibinfo {volume} {98}},\ \bibinfo {pages} {085101} (\bibinfo {year} {2018})}\BibitemShut {NoStop}%
\bibitem [{\citenamefont {Liu}\ \emph {et~al.}(2024{\natexlab{b}})\citenamefont {Liu}, \citenamefont {Zhao},\ and\ \citenamefont {Xiang}}]{LiuXiang_prb2024_geometricexcitations}%
  \BibitemOpen
  \bibfield  {author} {\bibinfo {author} {\bibfnamefont {Y.}~\bibnamefont {Liu}}, \bibinfo {author} {\bibfnamefont {T.}~\bibnamefont {Zhao}},\ and\ \bibinfo {author} {\bibfnamefont {T.}~\bibnamefont {Xiang}},\ }\bibfield  {title} {\bibinfo {title} {Resolving geometric excitations of fractional quantum Hall states},\ }\href {https://doi.org/10.1103/PhysRevB.110.195137} {\bibfield  {journal} {\bibinfo  {journal} {Phys. Rev. B}\ }\textbf {\bibinfo {volume} {110}},\ \bibinfo {pages} {195137} (\bibinfo {year} {2024}{\natexlab{b}})}\BibitemShut {NoStop}%
\bibitem [{\citenamefont {{Wang}}\ \emph {et~al.}(2025)\citenamefont {{Wang}}, \citenamefont {{Huxford}}, \citenamefont {{Nguyen}}, \citenamefont {{Ji}}, \citenamefont {{Kim}},\ and\ \citenamefont {{Yang}}}]{wangDynamics2025}%
  \BibitemOpen
  \bibfield  {author} {\bibinfo {author} {\bibfnamefont {Y.}~\bibnamefont {{Wang}}}, \bibinfo {author} {\bibfnamefont {J.}~\bibnamefont {{Huxford}}}, \bibinfo {author} {\bibfnamefont {D.~X.}\ \bibnamefont {{Nguyen}}}, \bibinfo {author} {\bibfnamefont {G.}~\bibnamefont {{Ji}}}, \bibinfo {author} {\bibfnamefont {Y.~B.}\ \bibnamefont {{Kim}}},\ and\ \bibinfo {author} {\bibfnamefont {B.}~\bibnamefont {{Yang}}},\ }\bibfield  {title} {\bibinfo {title} {{Dynamics and lifetime of geometric excitations in moir{\'e} systems}},\ }\href {https://doi.org/10.48550/arXiv.2502.02640} {\bibfield  {journal} {\bibinfo  {journal} {arXiv e-prints}\ ,\ \bibinfo {eid} {arXiv:2502.02640}} (\bibinfo {year} {2025})}\BibitemShut {NoStop}%
\bibitem [{\citenamefont {Wei\ss{}e}\ \emph {et~al.}(2006)\citenamefont {Wei\ss{}e}, \citenamefont {Wellein}, \citenamefont {Alvermann},\ and\ \citenamefont {Fehske}}]{Weisse_rmp2008_kpm}%
  \BibitemOpen
  \bibfield  {author} {\bibinfo {author} {\bibfnamefont {A.}~\bibnamefont {Wei\ss{}e}}, \bibinfo {author} {\bibfnamefont {G.}~\bibnamefont {Wellein}}, \bibinfo {author} {\bibfnamefont {A.}~\bibnamefont {Alvermann}},\ and\ \bibinfo {author} {\bibfnamefont {H.}~\bibnamefont {Fehske}},\ }\bibfield  {title} {\bibinfo {title} {The kernel polynomial method},\ }\href {https://doi.org/10.1103/RevModPhys.78.275} {\bibfield  {journal} {\bibinfo  {journal} {Rev. Mod. Phys.}\ }\textbf {\bibinfo {volume} {78}},\ \bibinfo {pages} {275} (\bibinfo {year} {2006})}\BibitemShut {NoStop}%
\bibitem [{\citenamefont {Pavarini}\ \emph {et~al.}(2019)\citenamefont {Pavarini}, \citenamefont {Koch},\ and\ \citenamefont {Zhang}}]{ED_spectralfunctions_koch}%
  \BibitemOpen
  \bibfield  {author} {\bibinfo {author} {\bibfnamefont {E.}~\bibnamefont {Pavarini}}, \bibinfo {author} {\bibfnamefont {E.}~\bibnamefont {Koch}},\ and\ \bibinfo {author} {\bibfnamefont {S.}~\bibnamefont {Zhang}},\ }\bibfield  {title} {\bibinfo {title} {Many-Body Methods for Real Materials : Autumn School organized by the Institute for Advanced Simulation at Forschungszentrum J\"{u}lich, 16 - 20 September 2019 : Lecture Notes of the Autumn School on Correlated Electrons 2019},\ }\href {https://doi.org/10.18154/RWTH-2019-09732} {\bibfield  {journal} {\bibinfo  {journal} {Autumn School on Correlated Electrons: Many-Body Methods for Real Materials}\ }\textbf {\bibinfo {volume} {J\"{u}lich}},\ \bibinfo {pages} {Diagramme (2019).} (\bibinfo {year} {2019})}\BibitemShut {NoStop}%
\bibitem [{\citenamefont {Lu}\ \emph {et~al.}(2024)\citenamefont {Lu}, \citenamefont {Chen}, \citenamefont {Wu}, \citenamefont {Sun},\ and\ \citenamefont {Meng}}]{Hongyu_prl2024_thermodinamicFCI}%
  \BibitemOpen
  \bibfield  {author} {\bibinfo {author} {\bibfnamefont {H.}~\bibnamefont {Lu}}, \bibinfo {author} {\bibfnamefont {B.-B.}\ \bibnamefont {Chen}}, \bibinfo {author} {\bibfnamefont {H.-Q.}\ \bibnamefont {Wu}}, \bibinfo {author} {\bibfnamefont {K.}~\bibnamefont {Sun}},\ and\ \bibinfo {author} {\bibfnamefont {Z.~Y.}\ \bibnamefont {Meng}},\ }\bibfield  {title} {\bibinfo {title} {Thermodynamic Response and Neutral Excitations in Integer and Fractional Quantum Anomalous Hall States Emerging from Correlated Flat Bands},\ }\href {https://doi.org/10.1103/PhysRevLett.132.236502} {\bibfield  {journal} {\bibinfo  {journal} {Phys. Rev. Lett.}\ }\textbf {\bibinfo {volume} {132}},\ \bibinfo {pages} {236502} (\bibinfo {year} {2024})}\BibitemShut {NoStop}%
\bibitem [{\citenamefont {M\"oller}\ \emph {et~al.}(2011)\citenamefont {M\"oller}, \citenamefont {W\'ojs},\ and\ \citenamefont {Cooper}}]{CooperMoller_prl2011_neutralfermion}%
  \BibitemOpen
  \bibfield  {author} {\bibinfo {author} {\bibfnamefont {G.}~\bibnamefont {M\"oller}}, \bibinfo {author} {\bibfnamefont {A.}~\bibnamefont {W\'ojs}},\ and\ \bibinfo {author} {\bibfnamefont {N.~R.}\ \bibnamefont {Cooper}},\ }\bibfield  {title} {\bibinfo {title} {Neutral Fermion Excitations in the Moore-Read State at Filling Factor $\ensuremath{\nu}=5/2$},\ }\href {https://doi.org/10.1103/PhysRevLett.107.036803} {\bibfield  {journal} {\bibinfo  {journal} {Phys. Rev. Lett.}\ }\textbf {\bibinfo {volume} {107}},\ \bibinfo {pages} {036803} (\bibinfo {year} {2011})}\BibitemShut {NoStop}%
\bibitem [{\citenamefont {Bernevig}\ and\ \citenamefont {Regnault}(2012)}]{Bernevig2012Emergent}%
  \BibitemOpen
  \bibfield  {author} {\bibinfo {author} {\bibfnamefont {B.~A.}\ \bibnamefont {Bernevig}}\ and\ \bibinfo {author} {\bibfnamefont {N.}~\bibnamefont {Regnault}},\ }\bibfield  {title} {\bibinfo {title} {Emergent many-body translational symmetries of Abelian and non-Abelian fractionally filled topological insulators},\ }\href {https://doi.org/10.1103/PhysRevB.85.075128} {\bibfield  {journal} {\bibinfo  {journal} {Phys. Rev. B}\ }\textbf {\bibinfo {volume} {85}},\ \bibinfo {pages} {075128} (\bibinfo {year} {2012})}\BibitemShut {NoStop}%
\bibitem [{\citenamefont {Pu}\ \emph {et~al.}(2023)\citenamefont {Pu}, \citenamefont {Balram}, \citenamefont {Fremling}, \citenamefont {Gromov},\ and\ \citenamefont {Papi\ifmmode~\acute{c}\else \'{c}\fi{}}}]{PuPapic_prl2023_moorereadsupersymmetry}%
  \BibitemOpen
  \bibfield  {author} {\bibinfo {author} {\bibfnamefont {S.}~\bibnamefont {Pu}}, \bibinfo {author} {\bibfnamefont {A.~C.}\ \bibnamefont {Balram}}, \bibinfo {author} {\bibfnamefont {M.}~\bibnamefont {Fremling}}, \bibinfo {author} {\bibfnamefont {A.}~\bibnamefont {Gromov}},\ and\ \bibinfo {author} {\bibfnamefont {Z.}~\bibnamefont {Papi\ifmmode~\acute{c}\else \'{c}\fi{}}},\ }\bibfield  {title} {\bibinfo {title} {Signatures of Supersymmetry in the $\ensuremath{\nu}=5/2$ Fractional Quantum Hall Effect},\ }\href {https://doi.org/10.1103/PhysRevLett.130.176501} {\bibfield  {journal} {\bibinfo  {journal} {Phys. Rev. Lett.}\ }\textbf {\bibinfo {volume} {130}},\ \bibinfo {pages} {176501} (\bibinfo {year} {2023})}\BibitemShut {NoStop}%
\bibitem [{hpc()}]{hpc2021}%
  \BibitemOpen
  \href {https://hpc.hku.hk/hpc/hpc2021/} {\bibinfo  {journal} {HPC2021, Information Technology Services, The University of Hong Kong}\ }\BibitemShut {NoStop}%
\bibitem [{par()}]{paratera}%
  \BibitemOpen
\bibfield  {journal} {  }\href {https://cloud.paratera.com} {\bibinfo  {journal} {Beijing PARATERA Tech CO.,Ltd}\ }\BibitemShut {NoStop}%
\bibitem [{\citenamefont {Haegeman}\ \emph {et~al.}(2011{\natexlab{b}})\citenamefont {Haegeman}, \citenamefont {Cirac}, \citenamefont {Osborne}, \citenamefont {Pi\ifmmode~\check{z}\else \v{z}\fi{}orn}, \citenamefont {Verschelde},\ and\ \citenamefont {Verstraete}}]{Haegeman2011}%
  \BibitemOpen
\bibfield  {journal} {  }\bibfield  {author} {\bibinfo {author} {\bibfnamefont {J.}~\bibnamefont {Haegeman}}, \bibinfo {author} {\bibfnamefont {J.~I.}\ \bibnamefont {Cirac}}, \bibinfo {author} {\bibfnamefont {T.~J.}\ \bibnamefont {Osborne}}, \bibinfo {author} {\bibfnamefont {I.}~\bibnamefont {Pi\ifmmode~\check{z}\else \v{z}\fi{}orn}}, \bibinfo {author} {\bibfnamefont {H.}~\bibnamefont {Verschelde}},\ and\ \bibinfo {author} {\bibfnamefont {F.}~\bibnamefont {Verstraete}},\ }\bibfield  {title} {\bibinfo {title} {{Time-Dependent Variational Principle for Quantum Lattices}},\ }\href {https://doi.org/10.1103/PhysRevLett.107.070601} {\bibfield  {journal} {\bibinfo  {journal} {Phys. Rev. Lett.}\ }\textbf {\bibinfo {volume} {107}},\ \bibinfo {pages} {070601} (\bibinfo {year} {2011}{\natexlab{b}})}\BibitemShut {NoStop}%
\bibitem [{\citenamefont {Haegeman}\ \emph {et~al.}(2016{\natexlab{b}})\citenamefont {Haegeman}, \citenamefont {Lubich}, \citenamefont {Oseledets}, \citenamefont {Vandereycken},\ and\ \citenamefont {Verstraete}}]{Haegeman2016}%
  \BibitemOpen
  \bibfield  {author} {\bibinfo {author} {\bibfnamefont {J.}~\bibnamefont {Haegeman}}, \bibinfo {author} {\bibfnamefont {C.}~\bibnamefont {Lubich}}, \bibinfo {author} {\bibfnamefont {I.}~\bibnamefont {Oseledets}}, \bibinfo {author} {\bibfnamefont {B.}~\bibnamefont {Vandereycken}},\ and\ \bibinfo {author} {\bibfnamefont {F.}~\bibnamefont {Verstraete}},\ }\bibfield  {title} {\bibinfo {title} {{Unifying time evolution and optimization with matrix product states}},\ }\href {https://doi.org/10.1103/PhysRevB.94.165116} {\bibfield  {journal} {\bibinfo  {journal} {Phys. Rev. B}\ }\textbf {\bibinfo {volume} {94}},\ \bibinfo {pages} {165116} (\bibinfo {year} {2016}{\natexlab{b}})}\BibitemShut {NoStop}%
\bibitem [{\citenamefont {Repellin}\ \emph {et~al.}(2014)\citenamefont {Repellin}, \citenamefont {Neupert}, \citenamefont {Papi\ifmmode~\acute{c}\else \'{c}\fi{}},\ and\ \citenamefont {Regnault}}]{repellin2014}%
  \BibitemOpen
  \bibfield  {author} {\bibinfo {author} {\bibfnamefont {C.}~\bibnamefont {Repellin}}, \bibinfo {author} {\bibfnamefont {T.}~\bibnamefont {Neupert}}, \bibinfo {author} {\bibfnamefont {Z.}~\bibnamefont {Papi\ifmmode~\acute{c}\else \'{c}\fi{}}},\ and\ \bibinfo {author} {\bibfnamefont {N.}~\bibnamefont {Regnault}},\ }\bibfield  {title} {\bibinfo {title} {{Single-mode approximation for fractional Chern insulators and the fractional quantum Hall effect on the torus}},\ }\href {https://doi.org/10.1103/PhysRevB.90.045114} {\bibfield  {journal} {\bibinfo  {journal} {Phys. Rev. B}\ }\textbf {\bibinfo {volume} {90}},\ \bibinfo {pages} {045114} (\bibinfo {year} {2014})}\BibitemShut {NoStop}%
\end{thebibliography}%

\appendix
\newpage\clearpage
\renewcommand{\theequation}{S\arabic{equation}} \renewcommand{\thefigure}{S%
	\arabic{figure}} \setcounter{equation}{0} \setcounter{figure}{0}

\begin{widetext}
\begin{center}
    \textbf{\Large Supplemental Material for}\\[0.5em]
    \textbf{\Large"Chiral Graviton Modes in Non-Abelian lattice Fractional Quantum Hall states"}\\[1em]
\end{center}
In this supplemental material, we provide additional data on the diagnosis of the Moore-Read ground state, bond dimension scale for TDVP result, and disk simulation.

\section*{Moore Read Ground state}
\subsection*{topological sector}
\begin{figure}[htp!]
    \centering
    \includegraphics[width=0.8\linewidth]{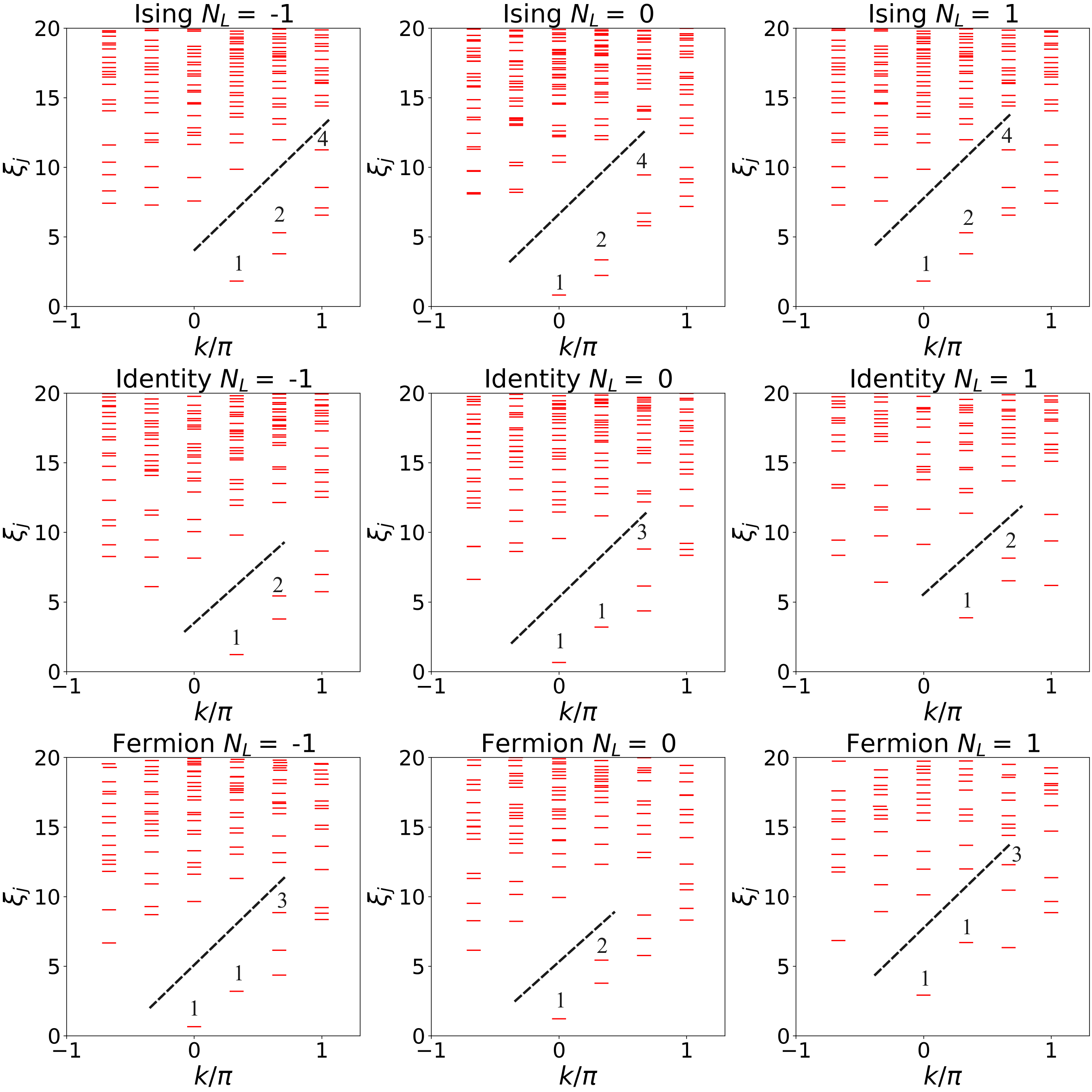}
    \caption{Entanglement diagnosis of Ising(top), Identity(middle) and Fermion(bottom) ground state. We show the momentum resolved entanglement spectrum for the charge sector with the highest weight ($N_L = 0$) and nearby $\pm 1$ sectors obtained from $L_y = 6$ cylinder.}
    \label{fig:supple_cft}
\end{figure}

In this section, we demonstrate the ground state diagnosis of MR states. The Entanglement Spectrum for the Ising, Identity, and Fermion ground state is shown in the top, middle, and bottom panels of Fig.~\ref{fig:supple_cft}. The Characteristic edge-mode counting $\{1,2,4,... \}$ for Ising sector and $\{1,1,3... \}$ for Identity$\&$Fermion is clearly observed. The even-odd difference in charge sector for Identity and Fermion sector is also seen. 

The IDMRG calculation for this result is performed in $L_y = 6$ cylinder with $1/6$ flux at $U = 0,V = 0$ case. The Graviton spectrum shown in Fig.1(d) is computed on top of this IDMRG result.

In the following section, we will also show that the finite descendant series are caused by finite $L_y$. 

\subsection{$L_y$ extrapolation}
The MPS results shown in the main text are computed on $L_y = 6$ cylinder. In this section, we provide more DMRG results on the MR ground state. The descendant series counting, as well as charge homogeneity, improves for larger circumference $L_y = 8$.

\begin{figure}[htp!]
    \centering
    \includegraphics[width=0.8\linewidth]{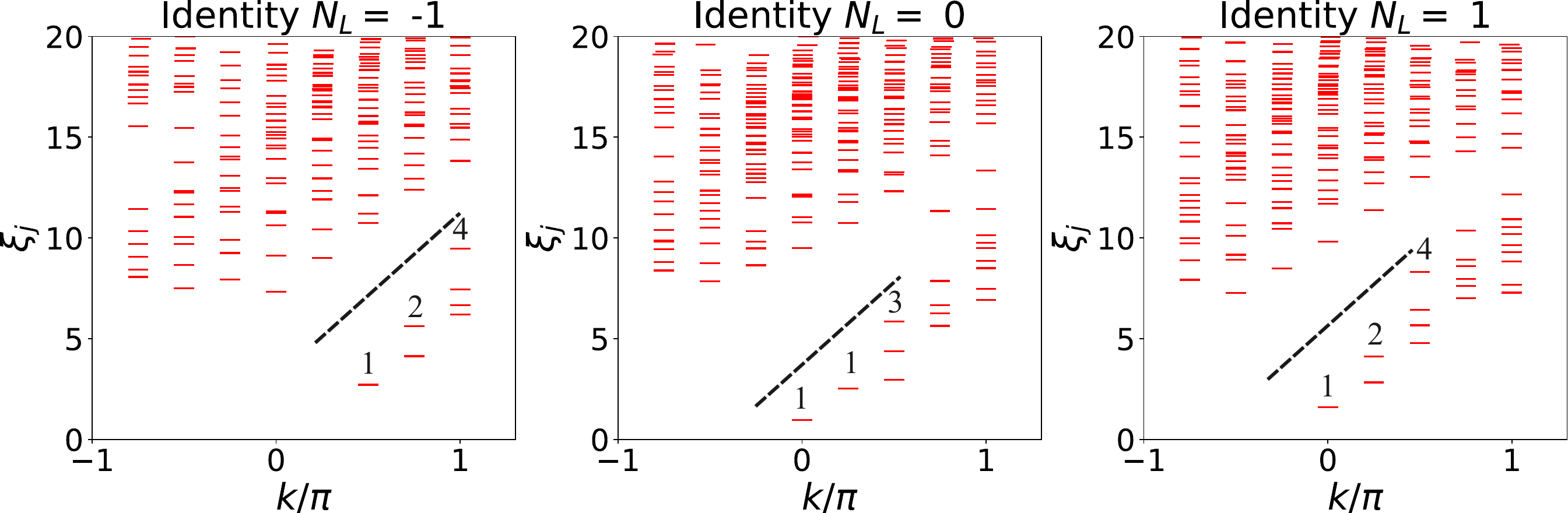}
    \caption{Entanglement Spectrum for Identity sector for the charge sector with the highest weight ($N_L = 0$) and nearby $\pm 1$ sectors, obtained from $L_y = 8$ cylinder.}
    \label{fig:supple_cftLy8}
\end{figure}

In Fig.~\ref{fig:supple_cftLy8}, we show the Entanglement spectrum for the identity sector in an $L_y = 8$ cylinder. The Characteristic edge-mode counting $\{1,2,4,... \}$ for odd charge sector and $\{1,1,3,... \}$ for even charge sector is observed. The length of the descendant series improves as we enlarge the circumference of the cylinder.

In addition, as shown in Fig.~\ref{fig:supple_chargeLy678}, the charge density $n(x) = \sum_y n_{x,y}/L_y$ also shows an oscillation behavior at $L_y = 6$, and this oscillation behavior gradually becomes weaker at larger $L_y$. 

We note that the TDVP calculation on $L_y$ exceeds our computational capacity, since the time evolution only converges after $D = 600$ (as shown in the following section); therefore, we stick to the $L_y = 6$ case.

\begin{figure}[htp!]
    \centering
    \includegraphics[width=0.4\linewidth]{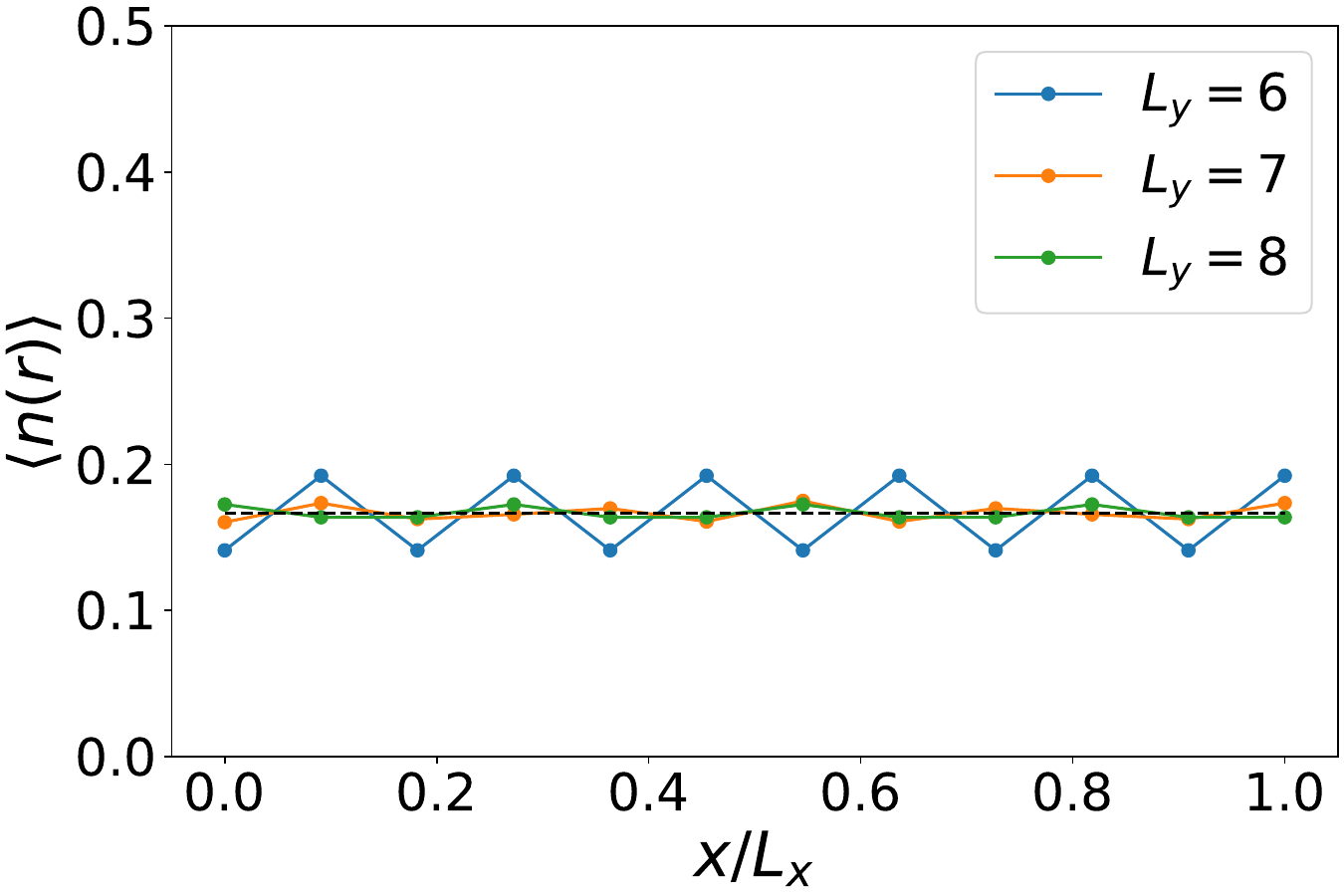}
    \caption{Charge density $n(x)$ along the cylinder for different $L_y$}
    \label{fig:supple_chargeLy678}
\end{figure}

\section*{TDVP Convergence check}
In MPS simulations, the spectrum is obtained via the Fourier transform of such a time-ordered Green's function:
\begin{equation}
\begin{aligned}
         I_{3b}^\pm (\omega) = \frac{1}{\mathcal{N}_-}  \int_0^T \ e^{i(\omega+i\gamma) t } \big( \langle O_{3b}^\mp(t)O_{3b}^\pm\rangle 
        -\langle O_{3b}^\pm \rangle \langle O_{3b}^\mp\rangle\big)dt.
\end{aligned}
\end{equation}
The bracket $\langle \bullet \rangle$ denotes the expectation value on the ground state. To compute the time-ordered Green's function, we first obtain the ground state via DMRG~\cite{White1992,White1993}, then time evolve $O^\pm |G\rangle$ using TDVP~\cite{Haegeman2011,Haegeman2016}. Note here that the resolution of the spectra is not only controlled by $\gamma$ but also by the total evolution time $T$. 


We measure the Green's function using TDVP, which projects the wave function at the next time slice $e^{i\hat{H}\Delta t}|\psi(t)\rangle$ to the tangent space of the MPS description of $|\psi(t)\rangle$. The numerical error depends on the bond dimension of the MPS, as MPS with larger D span a larger tangent space, thereby decreasing the projection error.

We check the convergence of our result on the MPS bond dimension $D$. In Fig~\ref{fig:supple_cylindertdvp_converg}, we show the real-time domain and frequency domain of the graviton response. A numerical artificial peak appears on the left panel for $D = 400$ and disappears when we enlarge to $D = 600$. The result is robust against further increase to $D = 800$, indicating the convergence of our result.

    
    

\begin{figure}[htp!]
    \centering
        \centering
        \includegraphics[width=0.8\linewidth]{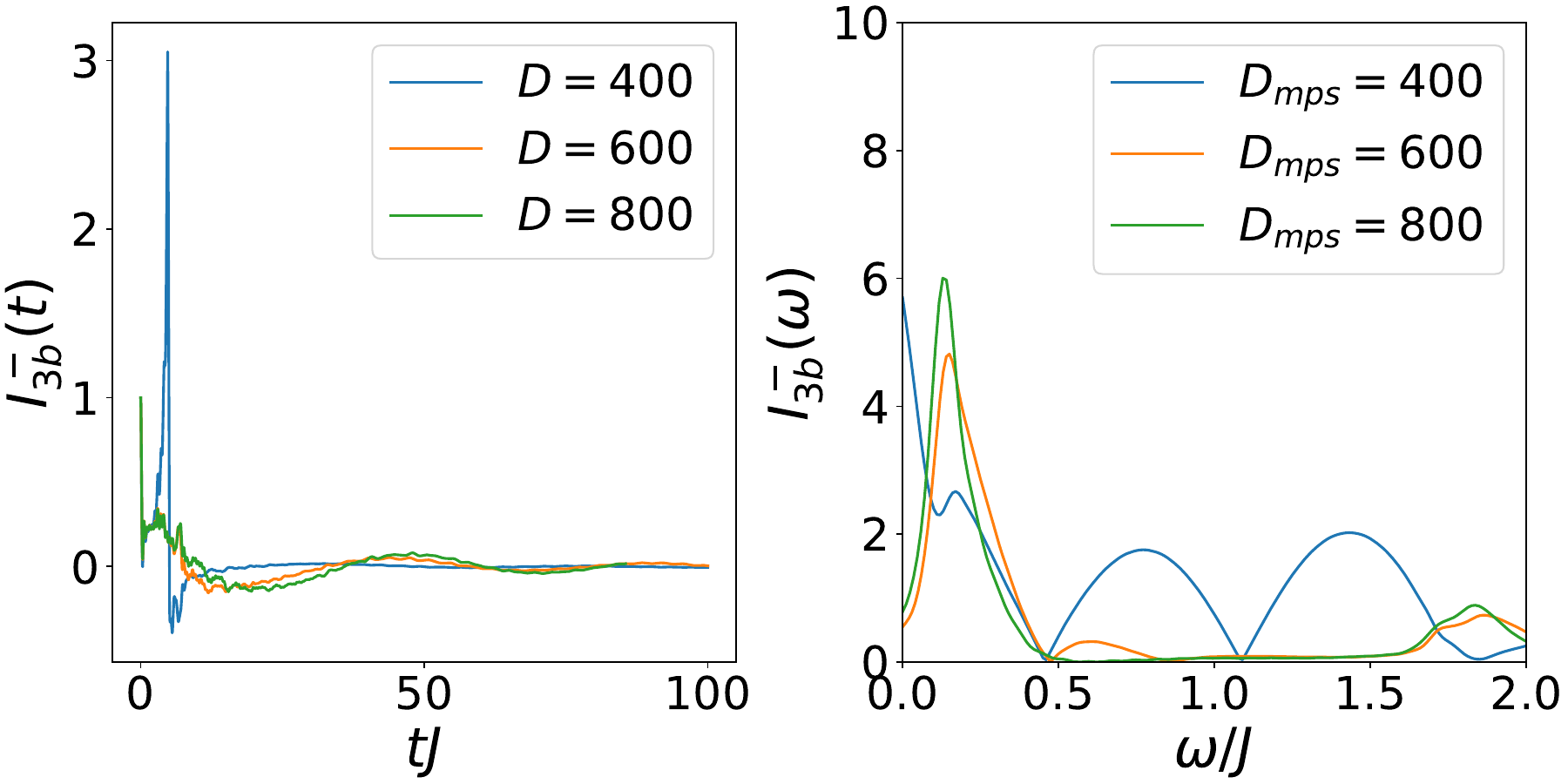}
        \caption{MPS convergence check on cylinder. (Left). Real-time correlation function for different bond dimensions. (Right), Fourier-transformed data for those in the left panel.}
    \label{fig:supple_cylindertdvp_converg}
    \vspace{2ex} 
        \centering
        \includegraphics[width=0.8\linewidth]{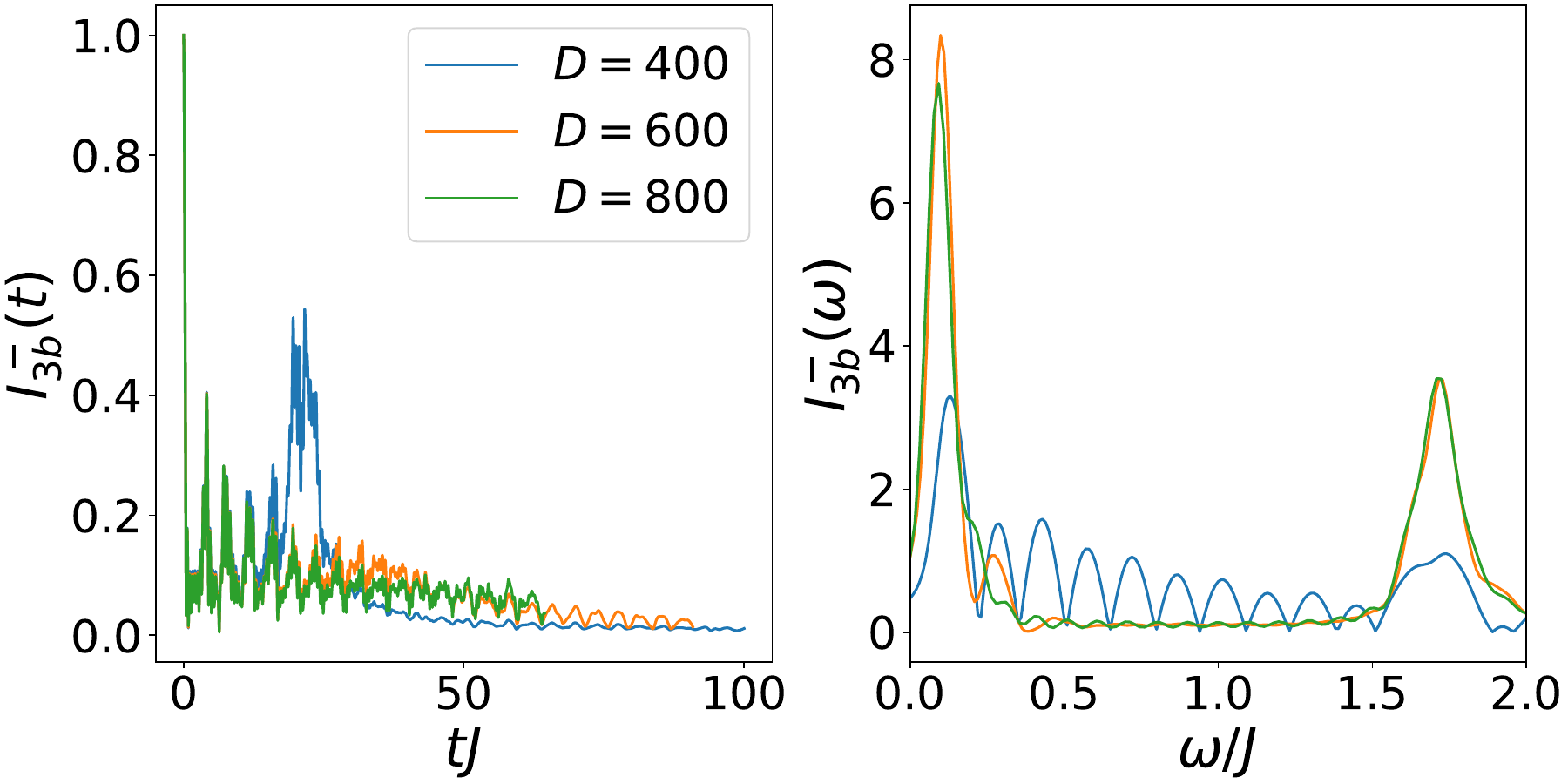}
        \caption{MPS convergence check on disk. (Left). Real-time correlation function for different bond dimensions. (Right), Fourier-transformed data for those in the left panel.}
    \label{fig:supple_Disktdvp_converg}
\end{figure}


We also check the convergence of the graviton spectrum on the disk (shown in Fig.~\ref{fig:supple_Disktdvp_converg}). The artificial peak also appears at $D = 400$ and disappears at larger $D$. The robustness of the graviton spectrum from $D = 600$ and $D = 800$ indicates the convergence.


\section*{Additional data on Disk simulation}
In this section, we provide the ground state charge density,2-body and 3-body correlation, ground state fidelity, quench comparison between even and odd numbers of particles, the effect of finite onsite repulsion $U$ on the graviton, and the evolution of correlation after quench.   

\begin{figure}[htp!]
    \centering
    \includegraphics[width=0.8\linewidth]{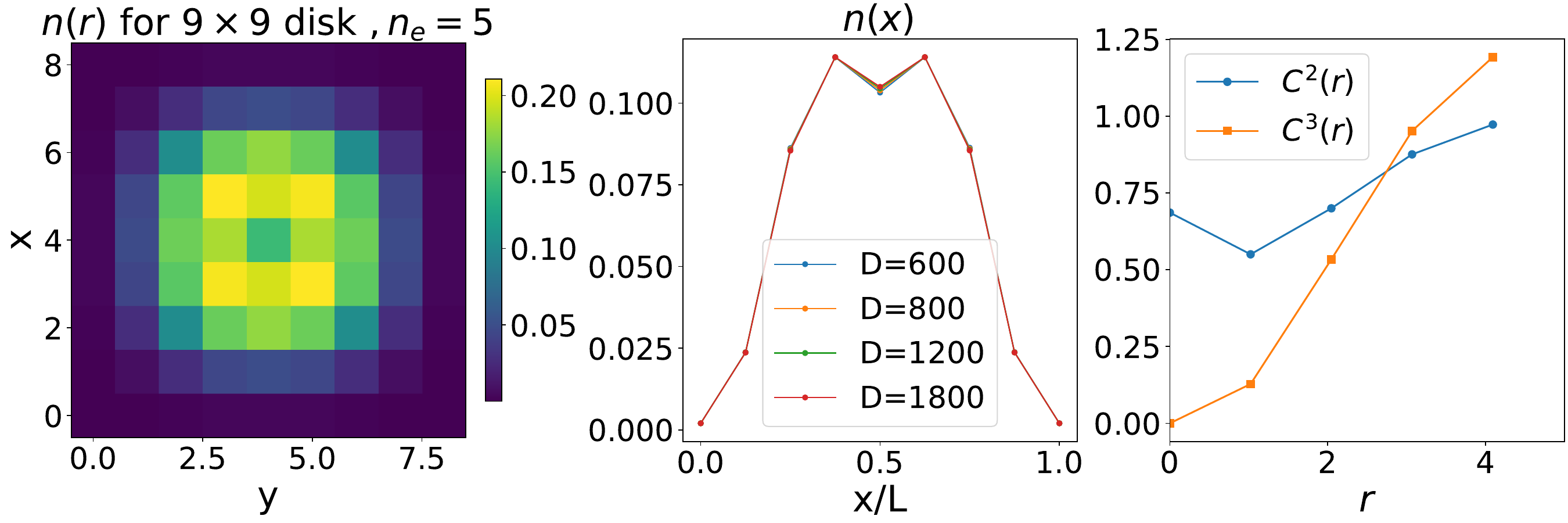}
    \caption{(Left).Ground state charge density. (Middle). $y-$direction averaged change density. (Right). 2-body and 3-body correlation.}
    \label{fig:supple_disk_gs}
\end{figure}

In Fig.~\ref{fig:supple_disk_gs}, we show the ground-state charge density distribution in the left and middle panels. The charge is confined to a ring that is offset from the initial point instead of localized in the center, a feature similar to those observed in experiment ~\cite{Joyce2026Pfaffian}. Consistent with our expectation, we observe compression of the 3-body correlation at short distances while the 2-body correlation remains finite, indicating the pairing structure of the Pfaffian state.

The ground state fidelity by tuning $\eta$ is shown in Fig.~\ref{fig:supple_disk_fidelity}. We choose the step $d\eta = 0.05$ and measure the ground state overlap $|\langle\psi(\eta)|\psi(\eta+d\eta)\rangle|$. The value is close to 1, indicating the robustness of the Paffian ground state in this $\eta$ window.
\begin{figure}[htp!]
    \centering
    \includegraphics[width=0.3\linewidth]{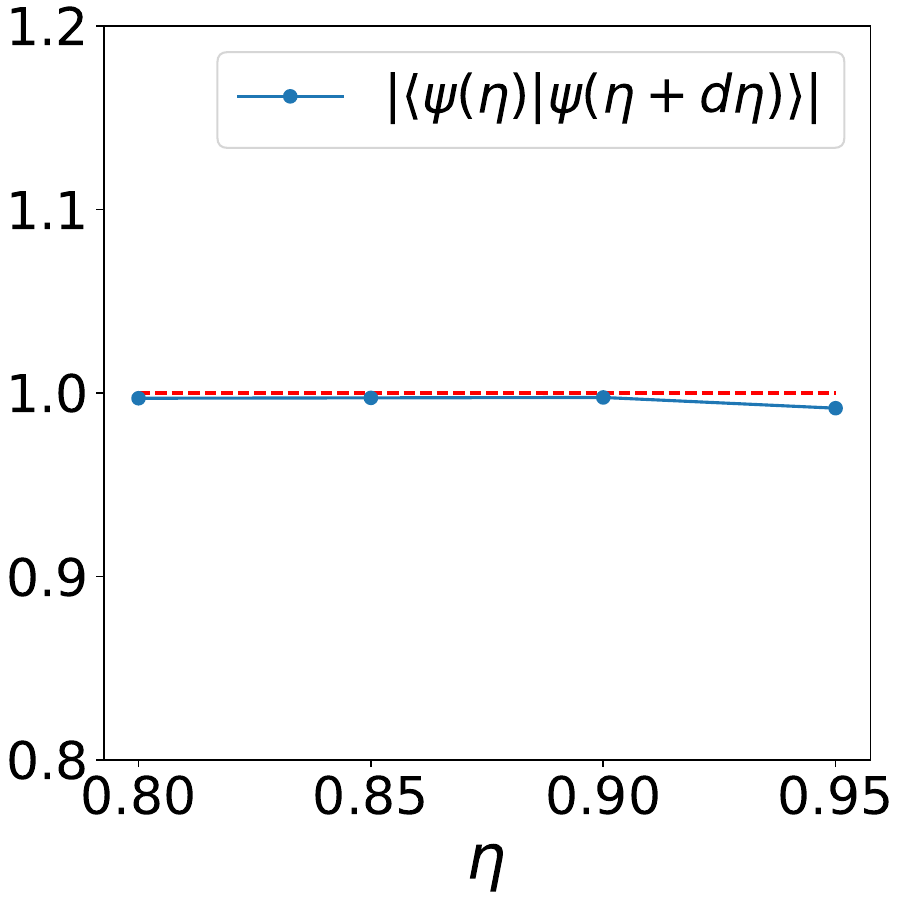}
    \caption{Ground state fidelity $|\langle\psi(\eta)|\psi(\eta+d\eta)\rangle|$. The red dashed line refers to the ideal fully overlapped case.}
    \label{fig:supple_disk_fidelity}
\end{figure}

\begin{figure}[htp!]
    \centering
    \includegraphics[width=0.4\linewidth]{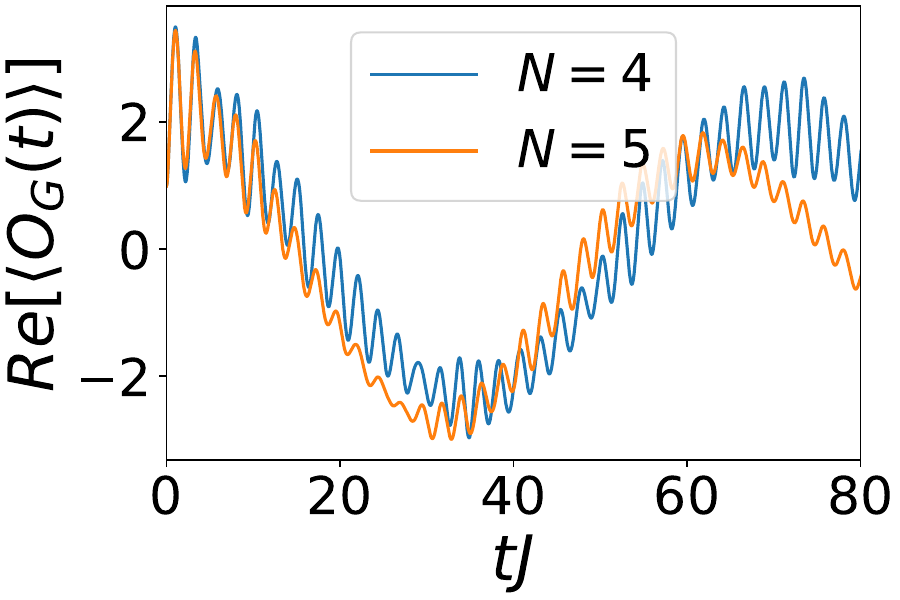}
    \caption{Quench simulation for even ($N = 4$) and odd ($N = 5$) number of particles. The disk size is $9\times9$ and $U = 0$, $V = 0$.}
    \label{fig:supple quench evenodd}
\end{figure}

An interesting aspect of the MR state is the even-odd effect, where in the case of an odd number of particles, the unpaired particles affect the neutral gap. We also simulate the quench dynamics for $N = 4$ particles. In Fig.~\ref{fig:supple quench evenodd}, we plot $Re[\langle O_{{G}}(t) \rangle]/Re[\langle O_{{G}}(0) \rangle]$; although the frequency of the oscillation changes a bit, the signal for both even and odd cases is clear. 

We also check the effect of on-site two-body repulsion terms $U\sum_i n_i (n_i-1)$ on the graviton spectrum. As in the experiment, a small $U$ term is needed to introduce a 3-body hard-core condition. However, adding extra interaction terms in the graviton spectrum calculation is not simply changing the Hamiltonian. The intrinsic metric tensor, thus the graviton operator we built, also changes. Therefore, the inclusion of a 2-body interaction term will bring 2-body terms in the expression of the graviton operator. 

We therefore introduce a weak $U = 0.1$ term and keep using the graviton operator we built in the main text. The result is shown in Fig.~\ref{fig:supple finite U graviton}; finite $U$ introduces a minor effect.

\begin{figure}[htp!]
    \centering
    \includegraphics[width=0.6\linewidth]{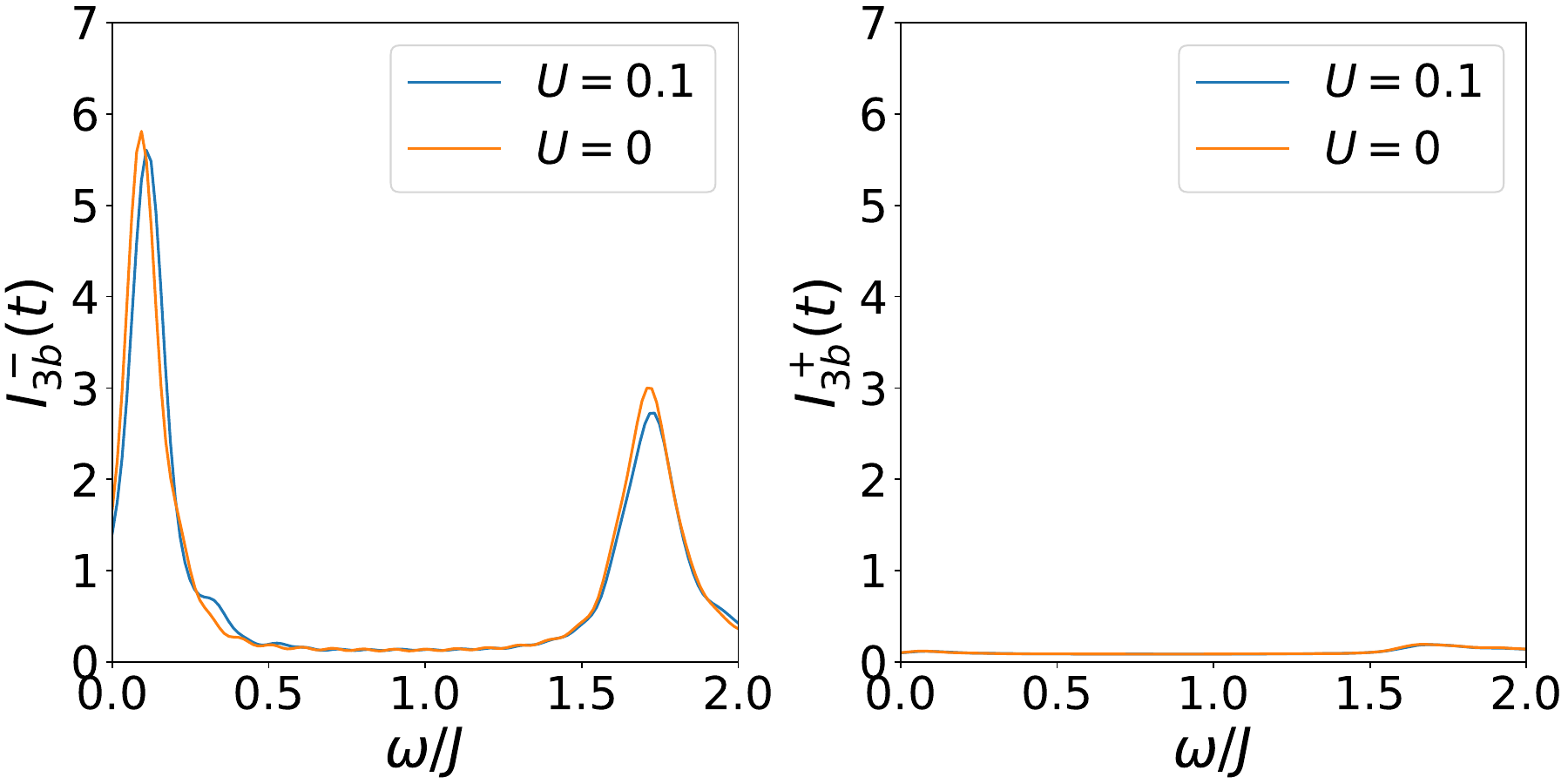}
    \caption{Graviton spectrum for $U = 0.1$ and $U = 0$. The calculation is on a  $9\times9$ disk and  $V = 0$.}
    \label{fig:supple finite U graviton}
\end{figure}

Finally, we present the dynamics of correlation holes after a quench in Fig~\ref{fig:supple quench_correlation hole}.  The snapshots reveal the counterclockwise
precession of the elliptical correlation hole, the correlation hole carries zero weight, the remaining part, which has a non-zero contribution to the dynamics, goes clockwise, and according to our convention of the positive direction of angle, the system thus possesses negative angular momentum.

Therefore, the chirality of the graviton mode could be determined by tracking the motion of the correlation hole.

\begin{figure}[htp!]
    \centering
    \includegraphics[width=0.4\linewidth]{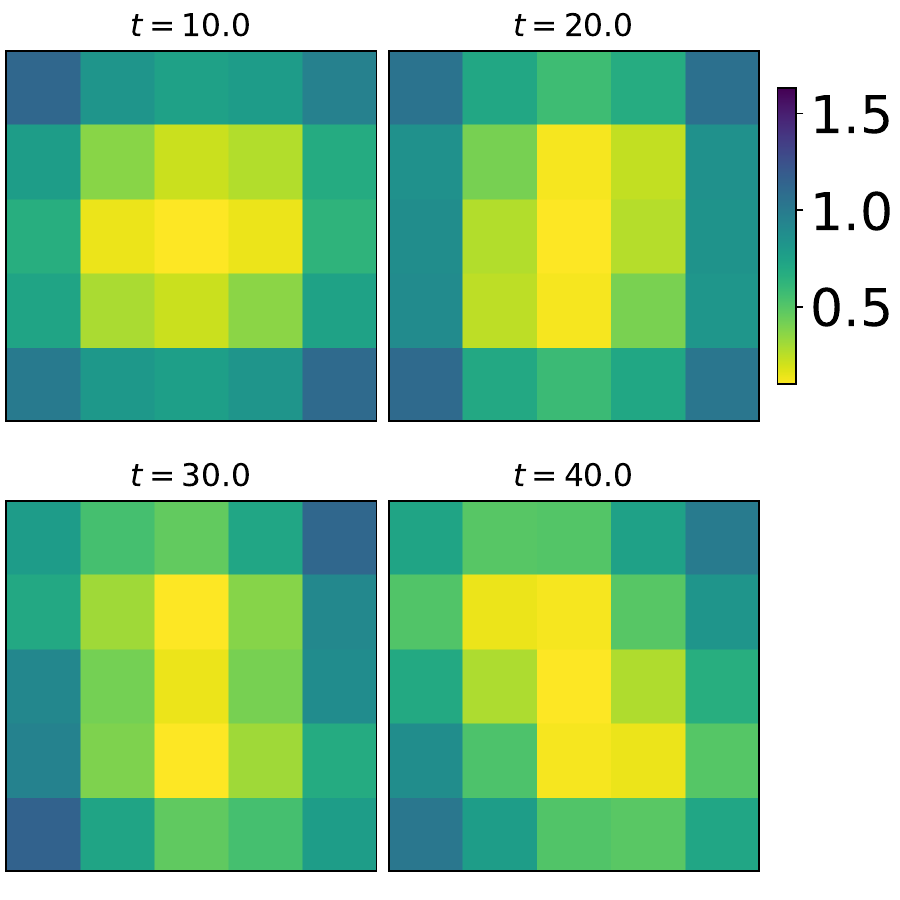}
    \caption{Snapshots of the correlation hole dynamics after a sudden quench. The plot is restricted to the center region of a $9 \times 9 $ disk. We start with a anistropic $\eta = 0.8$ ground state for $N = 4$ particles and quench it by changing to $\eta = 1$ at $t = 0$. Throughout the quench simulation, we set $U = 0$ and $V = 0$.}
    \label{fig:supple quench_correlation hole}
\end{figure}

\section*{Continuum limit}
We here take the continuum limit of the following operator:
\begin{align}
        O^\pm_{3b}=\sum_{ij}f(|r_i-r_j|)e^{\pm 2i\theta_{ij}} n_i(n_i-1) n_j
\end{align}
where $f(r)$ is a short range function that we shall keep unspecified at this point. The leading order for the three point function in the continuum is:
\begin{align}
   \lim_{a\ll l_B}  n_i(n_i-1) n_j = a^6 :\rho(r_i)\rho(r_i)\rho(r_i+\delta) :
\end{align}
where $::$ denotes normal ordering. Going to momentum space:
\begin{align}
    \rho_{\boldsymbol{q}}=\int d^2 r e^{i\boldsymbol{q}\cdot\boldsymbol{r}}\rho(\boldsymbol{r})
\end{align}
and performing a Lowest Landau level projection:
\begin{align}
   \rho_{\boldsymbol{q} }\rightarrow  \bar{\rho}_{\boldsymbol{q}}  e^{-\frac{1}{2}(ql_B)^2}
\end{align}
we arrive to the following expression:
\begin{align}
     O^\pm_{3b}\to \sum_{\boldsymbol{q}_1,\boldsymbol{q}_2,\boldsymbol{q}_3} \delta_{\boldsymbol{q}_1+\boldsymbol{q}_2+\boldsymbol{q}_3,0}  e^{-\frac{1}{2}(q_1^2+q_2^2+q_3^2)l_B^2} A_\pm(\boldsymbol{q}_3)  \Bar{\rho}_{\boldsymbol{q}_1} \Bar{\rho}_{\boldsymbol{q}_2} \Bar{\rho}_{\boldsymbol{q}_3}  
\end{align}
where:
\begin{align}
    A_\pm(\boldsymbol{q})= \int d^2 z   e^{i\boldsymbol{q}\cdot \boldsymbol{r}} e^{\pm 2i\theta_r} f(|\boldsymbol{r}|)
\end{align}
Now we can notice that because of the permutation symmetry in the integration domain we can also write:
\begin{align}
     O^\pm_{3n}\to \sum_{\boldsymbol{q}_1,\boldsymbol{q}_2,\boldsymbol{q}_3} \delta_{\boldsymbol{q}_1+\boldsymbol{q}_2+\boldsymbol{q}_3,0}  e^{-\frac{1}{2}(q_1^2+q_2^2+q_3^2)l_B^2} \left[A_\pm(\boldsymbol{q}_1)+A_\pm(\boldsymbol{q}_2)+A_\pm(\boldsymbol{q}_3)\right]\Bar{\rho}_{\boldsymbol{q}_1} \Bar{\rho}_{\boldsymbol{q}_2} \Bar{\rho}_{\boldsymbol{q}_3} 
\end{align}
As a last step we can follow the procedure we explained in Ref. \cite{longChiral2026} which, for a short range $f(r)$, leads to the expansion:
\begin{align}
    A_\pm(\boldsymbol{q}) \simeq (q_\pm)^2 (1+ O(|q|^2))
\end{align}
where $q_\pm =q_x\pm iq_y$. Feeding this expansion the previous expression and using the delta function to reduce one summation we get:
\begin{align}
     O^\pm_{3n}\to \sum_{\boldsymbol{q}_1,\boldsymbol{q}_2}   e^{-\frac{1}{2}(q_1^2+q_2^2+|\boldsymbol{q}_1+\boldsymbol{q_2}|^2)l_B^2}\; 2\left[ q_{1,\pm}^2 + q_{2,\pm}^2 + q_{1,\pm}q_{2,\pm}\right] \Bar{\rho}_{\boldsymbol{q}_1} \Bar{\rho}_{\boldsymbol{q}_2} \Bar{\rho}_{-\boldsymbol{q}_1-\boldsymbol{q}_2} 
\end{align}
The non-chiral expression reported in Ref. \cite{liou2019chiral} just correspond to a summation of the two chiralities.

\section{Details on neutral modes spectroscopy}

We here provide some more details on how to identify the two neutral modes from ED data on torus geometries. 

First of all, it is important to recall how this can be done in continuum Landaul Levels. Here there exist a center of mass symmetry which is not present on the lattice \cite{Bernevig2012Emergent} which can be used to label many-body eigenstates. In particular for $N$ particles filling a torus of size $L_x\times L_y=2\pi l_B^2 N$, due to relative and center of mass translations in:
\begin{align}
    BZ_{FQH}=\{ (2\pi n_x/L_x,2\pi n_y/L_y)\; \;\mathrm{with}\;\;n_{x/y}=0,...,N-1\}
\end{align}
giving a total of $N^2$ possible values. For the bosonic $\nu=1$ Moore-Read state, there are three (one) ground states for even (odd) number of particles on the torus. These in particular live in different center of mass momentum sectors $\{(0,0); (0,\pi N/L_y);(\pi N/L_x,0)\}$ for even $N$ and in $\{(0,0)\}$ for odd $N$. 

For the Harper hofstader models at flux $\phi=1/q$ and filling $\nu=1$ considered in the main text with size $L_x\times L_y$ (lattice constant $a=1$), we have a unit cell of size $1\times q$ (hence $L_y = m q$ and $L_x=N/m$). The lattice reduced momentum symmetries \cite{Bernevig2012Emergent}, and we are left with many-body lattice momenta which live in a Brillouin zone:
\begin{align}
    BZ_{HH}=\{ (2\pi n_x/L_x,2\pi n_y/L_y)\; \;\mathrm{with}\;\;n_{x}=0,...,N/m-1; n_y=0,...,m-1\}
\end{align}
which amounts to $N$ total sectors instead of the $N^2$ in the continuum. Indeed the continuum total momenta from $BZ_{FQH}$ are mapped onto $BZ_{HH}$, as the center of mass symmetry is not present on lattice but is only emergent \cite{Bernevig2012Emergent}. For a specific choice of system size $N\times q$, the momenta $(0,0)$ and $(\pi,0)$ are mapped into themselves. Note that from these two it is possible to reconstruct magnetoroton and neutral fermion dispersions \cite{Repellin2015projective}. Other momentum points $(0,\pi)$ and $(\pi,\pi)$ are instead mapped into others and defining dispersions around those do not produce nice dispersions as in Fig. 3 of the main text. There in particular we make use of \textit{unfolded} density operator with momenta $\boldsymbol{k}\in\{(2\pi n_x/ L_x,2\pi n_y/L_y) ; n_{x/y}=0,..,L_{x/y}-1\}$ that increase the effective momentum resolution of the operator and account for variations within the unit cell $1\times q$. Note that this construction effectively correspond to the single magnetoroton ansatz developed for abelian FCI \cite{repellin2014}. The fact that the neutral fermion mode does not have a large overlap with it, i.e. its matrix element are small, signals that its character is not captured by a bosonic density operator.

\section{Comparison full and projected ED}
We here compare projected ED results with full ED calculations to assess the strength of band mixing effects.

Projecting to the lowest band strongly reduces the complexity as one needs to simulate $N$ bosons in $N_\Phi=\phi L_x L_y$ lowest band orbitals rather than $N_s=L_x L_y$ sites. The presence of a band gap $\Delta_{b}$ allows for an expansion around the non-interacting limit $W=0$, with expected corrections of order $W^2/\Delta$ involving virtual population in higher band states. In Fig. \ref{fig:truncation_check}(a) we show the many-body gap in the 0 momentum sector as a function of $W$ for  both projected ED and full ED on the same $6\times 6$ torus with $\phi=1/6$. Band mixing effects start to kick in at around $W\sim 2$. At large $W$ the gap in full ED flattens out, eventually reaching a finite value in the $W\to \infty $ limit. In Fig. \ref{fig:truncation_check}(b) we further show the low-lying netural specturm resolved in momentum where it is possible to see how band mixing effects give an overall rescaling of the energy scales. As such, the precise energy of the graviton mode calculated in the main text shall not be taken as an exact value. Note however that as the graviton decay is mainly due to decay into many-body excitations which mostly live in the lowest band we expect the physics of the graviton lifetime to be well captured also in a band-projected calculation.
\begin{figure}
    \centering
    \begin{overpic}[width=0.5\linewidth]{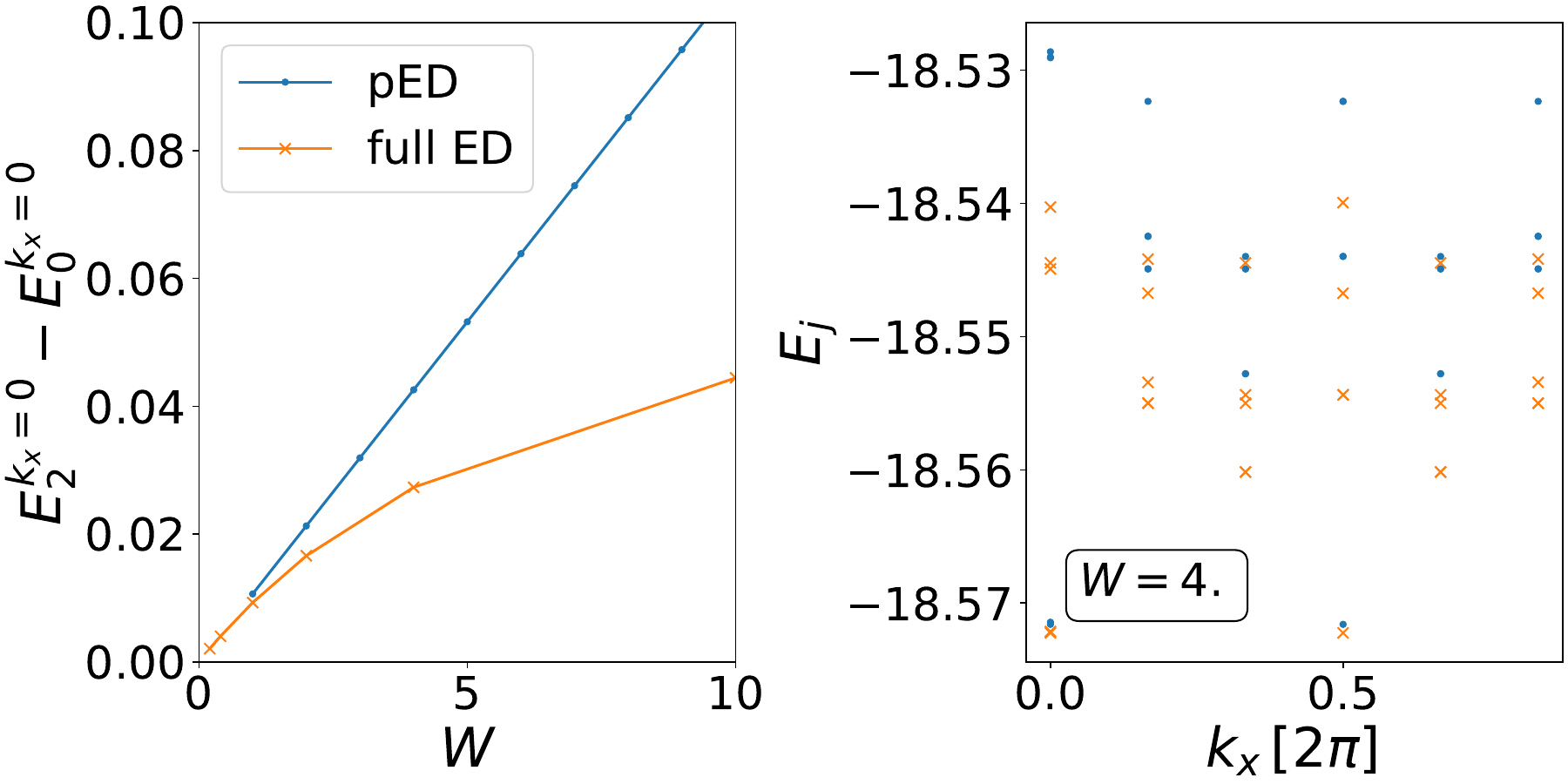}
    \put(90,10){(b)}
    \put(40,10){(a)}
    \end{overpic}
    \caption{Comparison of full and projected ED calculations at $\phi=1/6$ for a system of $N=6$ bosons in a $6\times 6$ torus.}
    \label{fig:truncation_check}
\end{figure}

\end{widetext}

\end{document}